\documentclass[a4paper,11pt]{article}
\usepackage[mathlines]{lineno}
\usepackage{jcappub}

\usepackage{hyperref}

\usepackage[table,svgnames,dvipsnames]{xcolor}
\usepackage{colortbl}

\usepackage[normalem]{ulem}
\usepackage{aas_macros}
\usepackage{subcaption}
\usepackage{graphicx}
\usepackage{graphics}
\usepackage{dcolumn}
\usepackage{bm}
\usepackage{orcidlink}
\usepackage{cases}
\usepackage{booktabs}
\usepackage{comment}
\usepackage{multirow}
\usepackage{makecell}
\usepackage{siunitx}
\usepackage{tabularx}
\usepackage{xspace}
\usepackage{soul} 
\graphicspath{{figs/}}
\usepackage{tensor}
\usepackage{amsmath}

\usepackage{listings}
\usepackage{caption}

\lstdefinelanguage{yaml}{
  keywords={true, false, null, yes, no, y, n},
  keywordstyle=\color{blue},
  basicstyle=\ttfamily\scriptsize,
  comment=[l]{\#},
  commentstyle=\color{gray},
  stringstyle=\color{orange},
  moredelim=[l][\color{brown}]{\&},
  moredelim=[l][\color{magenta}]{*},
  moredelim=[s][\color{gray}]{:}{\ },
  sensitive=false
}

\usepackage{orcidlink} 
\usepackage{hanging} 
\usepackage{arydshln}

\newcommand{\SNGATE}[1]{\textbf{\textcolor{red}{[SNIa issue:]}}}

\renewcommand{\arraystretch}{1.4}
\def\be{\begin{equation}}
\def\ee{\end{equation}}

\def\ba#1\ea{\begin{align*}#1\end{align*}}

\newcommand{\code}[1]{{\texttt{#1}}}
\renewcommand{\emph}[1]{\textit{#1}}
\definecolor{RoyalBlue}{rgb}{0.25,.41,.88}
\definecolor{WildStrawberry}{HTML}{EE2967}
\definecolor{RedWine}{rgb}{0.743,0,0}
\definecolor{bittersweet}{rgb}{1.0, 0.44, 0.37}
\definecolor{burntorange}{rgb}{0.8, 0.33, 0.0}
\definecolor{midnightgreen}{rgb}{0.0, 0.29, 0.33}
\definecolor{otherblue}{rgb}{0.20, 0.73, 0.92}

\usepackage[nameinlink,noabbrev]{cleveref}
\crefname{equation}{Eq.}{Eqs.}
\crefname{section}{Section}{Sections}
\crefname{figure}{Figure}{Figures}
\crefname{table}{Table}{Tables}
\crefname{appendix}{Appendix}{Appendices}
\Crefname{figure}{Figure}{Figures}
\Crefname{equation}{Equation}{Equations}
\Crefname{section}{Section}{Sections}
\Crefname{table}{Table}{Tables}

\newcommand{\mksym}[1]{\ifmmode {\rm #1}\else #1\fi}

\newcommand{\ob}{\omega_\mathrm{b}}
\newcommand{\ocdm}{\omega_\mathrm{cdm}}

\newcommand{\lcdm}{$\Lambda$CDM} 
 
\usepackage{soul}

\usepackage{bm}
\let\vec\bm

\newcommand{\hmpci}{\,h\text{Mpc}^{-1}}

\newcommand{\Dk}[1]{\frac{d^3#1}{(2\pi)^3}}

\newcommand{\vk}{\vec k}
\newcommand{\vp}{\vec p}

\title{Testing Scale-Dependent Modified Gravity with DESI DR1}

\emailAdd{aviles@icf.unam.mx}

\affiliation{Affiliations are in Appendix \ref{sec:affiliations}}

\author[1,2]{{D.~Gonzalez}\orcidlink{0009-0009-6485-640X},}
\author[1,3]{{G.~Niz}\orcidlink{0000-0002-1544-8946},}
\author[2,3]{{A.~Aviles}\orcidlink{0000-0001-5998-3986},}
\author[4,5]{{C.~Garcia-Quintero}\orcidlink{0000-0003-1481-4294},}
\author[2]{{H.~E.~Noriega}\orcidlink{0000-0002-3397-3998},}
\author[6]{{J.~Aguilar},}
\author[7]{{S.~Ahlen}\orcidlink{0000-0001-6098-7247},}
\author[8,9]{{D.~Bianchi}\orcidlink{0000-0001-9712-0006},}
\author[10]{{D.~Brooks},}
\author[6]{{T.~Claybaugh},}
\author[11]{{A.~de la Macorra}\orcidlink{0000-0002-1769-1640},}
\author[12]{{A.~de~Mattia}\orcidlink{0000-0003-0920-2947},}
\author[10]{{P.~Doel},}
\author[6,13]{{S.~Ferraro}\orcidlink{0000-0003-4992-7854},}
\author[14,15]{{J.~E.~Forero-Romero}\orcidlink{0000-0002-2890-3725},}
\author[16,17,18]{{E.~Gaztañaga}\orcidlink{0000-0001-9632-0815},}
\author[19]{{S.~{Gontcho A Gontcho}}\orcidlink{0000-0003-3142-233X},}
\author[20]{{G.~Gutierrez},}
\author[21]{{C.~Hahn}\orcidlink{0000-0003-1197-0902},}
\author[22,23,24]{{K.~Honscheid}\orcidlink{0000-0002-6550-2023},}
\author[25,26]{{D.~Huterer}\orcidlink{0000-0001-6558-0112},}
\author[27]{{M.~Ishak}\orcidlink{0000-0002-6024-466X},}
\author[28]{{R.~Joyce}\orcidlink{0000-0003-0201-5241},}
\author[28]{{S.~Juneau}\orcidlink{0000-0002-0000-2394},}
\author[29]{{R.~Kehoe},}
\author[30]{{D.~Kirkby}\orcidlink{0000-0002-8828-5463},}
\author[6]{{M.~Landriau}\orcidlink{0000-0003-1838-8528},}
\author[31]{{L.~Le~Guillou}\orcidlink{0000-0001-7178-8868},}
\author[6]{{M.~E.~Levi}\orcidlink{0000-0003-1887-1018},}
\author[32,33]{{M.~Manera}\orcidlink{0000-0003-4962-8934},}
\author[28]{{A.~Meisner}\orcidlink{0000-0002-1125-7384},}
\author[34,33]{{R.~Miquel},}
\author[17]{{S.~Nadathur}\orcidlink{0000-0001-9070-3102},}
\author[35,36,37]{{W.~J.~Percival}\orcidlink{0000-0002-0644-5727},}
\author[38]{{I.~P\'erez-R\`afols}\orcidlink{0000-0001-6979-0125},}
\author[39]{{G.~Rossi},}
\author[40,41,42]{{L.~Samushia}\orcidlink{0000-0002-1609-5687},}
\author[43]{{E.~Sanchez}\orcidlink{0000-0002-9646-8198},}
\author[6]{{D.~Schlegel},}
\author[44]{{H.~Seo}\orcidlink{0000-0002-6588-3508},}
\author[6]{{J.~Silber}\orcidlink{0000-0002-3461-0320},}
\author[28]{{D.~Sprayberry},}
\author[26]{{G.~Tarl\'{e}}\orcidlink{0000-0003-1704-0781},}
\author[28]{{B.~A.~Weaver},}
\author[6]{{R.~Zhou}\orcidlink{0000-0001-5381-4372},}

\date{\today}

\abstract{
The Dark Energy Spectroscopic Instrument (DESI) provides an unprecedented opportunity to test deviations from general relativity (GR) that introduce a new physical scale within its redshift range. Using the connection between a Yukawa-like potential and the Hu–Sawicki $f(R)$ model, we place strong constraints on the range of a hypothetical fifth force mediated by a massive scalar field. We analyze the power spectrum measurements from DESI Data Release 1 using a baseline EFT model that employs the \texttt{fkpt} approach for the loop integrals. We find no evidence for deviations from GR and obtain the constraint $\log_{10}|f_{R_0}| < -4.59$ (95\% C.L.), where $f_{R_0}\equiv df(R)/dR|_{\mathrm{today}}$. This corresponds to an upper bound at redshift zero on the scale at which corrections to GR become important, $\lambda < 17.81$ Mpc, or equivalently, a lower bound on the mass of the additional gravitational mediator of $m_\phi > 3.60 \times 10^{-31}$ eV. We find that the modified gravity parameter $f_{R_0}$ is largely orthogonal to the cosmological parameters in the model, such that no additional projection effects relative to the GR case are introduced in this Full-Shape analysis. Furthermore, a second modified gravity parameter, the power index $n$, which modulates the time-variation of the associated mass, is found to be consistent with previous analyses that fixed it to unity. Adding DESI BAO data or other cosmological probes does not significantly change these results. The conclusions remain similar if the background evolution is described by evolving dark energy instead of a cosmological constant. Additionally, we test the robustness of the baseline model by varying the maximum wavenumber used in the Full-Shape analysis and analyzing the DESI targets separately. Finally, we analyze the degeneracies between the modified-gravity parameters and the sum of neutrino masses.}

\begin{document}
\maketitle
\flushbottom

\section{Introduction and summary of main results}
\label{sec:intro}

The Dark Energy Spectroscopic Instrument (DESI), a wide-angle instrument in the Mayall telescope at Kitt Peak \cite{DESI2022.KP1.Instr}, Arizona, provides an unprecedented capability to test modifications to Einstein equations during the history of the universe \cite{DESI2016a.Science}, in particular using its most recent release of extra-galactic targets with accurate redshift measurements \cite{DESI2024.I.DR1}.
 
At the background level, the DESI Collaboration has presented hints of deviations from the standard cosmological model, $\Lambda$CDM. Using a time-dependent equation of state for the dark energy component in the DESI Data Release 2 (DR2) baryon acoustic oscillations (BAO) data combined with cosmic microwave background (CMB) and type Ia supernovae (SNIa), the most recent collaboration analysis exhibits moderate deviation with $\Lambda$CDM at more than 2.5$\sigma$ \cite{DESI.DR2.BAO.cosmo, Y3.cpe-s1.Lodha.2025}. Although new data, analysis techniques, and improved determination of systematic errors can shed light on the robustness of this departure from the standard model, if the signal is still present it would be implausible, from a pure background evolution, to interpret it as  modification to the gravitational theory, instead of a particle physics component which is minimally coupled to gravity. 

At the next order in a cosmological perturbation expansion, the DESI collaboration has also analyzed a two-parameter family of modifications to Einstein’s theory, which are constant or smoothly dependent on time \cite{DESI2024.VII.KP7B,KP7s1-MG}. This analysis shows that the DESI data are compatible with general relativity (GR).
However, a point to notice is that modified gravity (MG) generally introduces new characteristic length scales,
hence the importance to contrast this paradigm with observations. In this context, what characterizes DESI over other cosmological probes at the present time is its unique power to test alternative theories of gravity which introduce a new physical scale which leaves imprints on structure formation. 

Galaxy surveys are particularly well suited to detect a hypothetical additional physical scale introduced by MG, $k_\text{MG}^{-1}$. In many models that aim to explain cosmic acceleration without dark energy, deviations from GR arise below a characteristic scale, while clustering on larger scales remains essentially unchanged. As a result, the galaxy power spectrum is modified for wavenumbers $k \gtrsim k_\text{MG}$, and reduces to the GR prediction on the largest scales. Moreover, in these scenarios the range of the additional force is typically time dependent, becoming relevant only at late times. This limits the constraining power of probes such as the CMB,  which becomes affected by MG primarily through lensing only.   Furthermore, background probes such as SNIa or BAO are insensitive to these scale-dependent effects, emphasizing the role of galaxy clustering observables that capture the scale dependence of the power spectrum as a key tool to test such models.

Linearized Horndeski theories typically reduce to Newtonian gravity with a Yukawa-type correction; thus a natural way to study scale-dependent modified gravity is to introduce a new scalar gravitational degree of freedom that mediates a fifth force through a Yukawa potential,
\begin{equation}\label{eqn:yukawa}
V(r)\sim G_N\frac{2\beta^2}{r} e^{-m_{\phi}r},
\end{equation}
characterized by an additional physical scale $\lambda= \frac{1}{m_{\phi}}$ and a strength $2\beta^2$. In Fourier space, this scale corresponds to a characteristic comoving wavenumber $k_\text{MG} = a m_\phi$, which separates the regime where deviations from GR become relevant. In this context, $a$ denotes the scale factor, normalized to unity today, and related to the redshift by $a=(1+z)^{-1}$.  In our nomenclature, scale-dependent MG is bounded to cases where MG signatures only appear at ranges of the order and smaller than the characteristic length introduced by the theory, even though this quantity and the MG strength may vary with time. Furthermore, it should fall within the range explored by cosmological probes, in particular DESI, such that at very large scales the theory reduces to the background evolution in GR. In contrast, what we refer to scale-independent MG describes cases where the departure from Einstein equations is not only constant but also scale-dependent with the new scale existing outside the range of cosmological tests. The case of Dvali–Gabadadze–Porrati (DGP) gravity \cite{Dvali:2000hr} is an example of the latter option, where the transition from the brane-world scenario to a 4-dimensional metric gravity occurs at scales much larger than the Hubble scale. At the technical level, the key distinction between scale-dependent and scale-independent MG tests, is that in the former case one obtains the information from the whole shape of the power spectrum, expecting  enhancements for $k>k_\text{MG}$, while in the latter only from its amplitude and the growth rate. 

The models analyzed by the DESI collaboration mentioned above \cite{DESI2024.VII.KP7B,KP7s1-MG} fall within the scale-independent class and are therefore constrained mainly through the monopole-to-quadrupole ratio in the power spectrum at large scales, or equivalently the growth rate, $f$. In contrast, theories that converge to GR on very large scales but deviate at intermediate scales ($k/a > m_\phi$) are constrained by the detailed shape of the power spectrum, with minimal influence from the largest-scale modes or from background evolution. This constitutes a fundamentally different strategy for testing departures from GR, an advantage that is absent in the scale-independent approach.

In this work, we set limits on this new scale using DESI Full-Shape (FS) data for the first time.\footnote{A complementary analysis in \cite{Rauhut:2025eaz}, based on DESI DR1 and weak lensing surveys, also constrains mild scale-dependent modifications to general relativity using the $E_G$ estimator.
} To this end, we perform an analysis  of the power spectrum of discrete tracers in the DESI Data Release 1 (DR1), whose public description can be found in \cite{DESI2024.I.DR1}. This pipeline has been validated and many of the systematics, such as projection effects, have also been understood or accounted for \cite{DESI2024.V.KP5,KP5s3-Noriega,KP5s1-Maus,KP5s2-Maus}. 

The modeling of MG within an effective field theory (EFT) framework has been extensively studied in the literature (see, e.g., \cite{Koyama:2009me,Taruya:2013quf,Brax:2013fna,Bellini:2015oua,Taruya:2014faa,Taruya:2016jdt,Winther:2017jof,Fasiello:2017bot,Bose:2016qun,Aviles:2017aor,Bose:2017dtl,Cusin:2017wjg,Bose:2018orj,Valogiannis:2019xed,Valogiannis:2019nfz,Aviles:2018qot,Aviles:2018saf,Aviles:2020wme,Rodriguez-Meza:2023rga,Euclid:2023bgs,Aviles:2024zlw,Zheng:2025owb}). 
We adopt the approach of \cite{Aviles:2017aor,Aviles:2020wme,Rodriguez-Meza:2023rga} and utilize the \textsc{fkpt} code,\footnote{\url{https://github.com/alejandroaviles/fkpt}.}
 which we adapt for integration into the \textsc{Desilike} software package\footnote{\url{https://github.com/cosmodesi/desilike}.}
 to include perturbative kernels beyond Einstein-de Sitter, as is usual in $\Lambda$CDM - EFT approaches. In particular, we take advantage of the mapping between the Yukawa potential (\ref{eqn:yukawa}) and the well-known $f(R)$ Hu-Sawicki (HS) model \cite{Hu:2007nk}, when screening effects that restore GR on short scales are not considered.\footnote{We utilize a conservative range of scales, for which screening effects have been shown to be small and degenerate with EFT counterterms \cite{Rodriguez-Meza:2023rga}.
} A derivation of this mapping is given, e.g., in \cite{Amendola:2019laa}. The precise expression for the potential derived from HS is described later, but at this point it is sufficient to understand that the link relating both descriptions translates into a correspondence between the mass scale $m_\phi$ with the HS parameters ($n$, $f_{R_0}$) and the matter abundance today, $\Omega_{m}$, given by 
\begin{equation}\label{eq:HS-mass-relation}
    m^{2}_{\phi}(a)=\frac{H_0^{2}}{2|f_{R_0}|}\frac{(\Omega_{m}a^{-3}+4\Upsilon(a))^{n+2}}{(\Omega_{m}+4\Upsilon(a_0))^{n+1}},
\end{equation}
where we choose two different dark energy models for describing the background evolution: a pure cosmological constant with $\Upsilon(a)=(1-\Omega_{m})$, and the $wowa$-parametrization (also known as CPL \cite{Chevallier:2000qy, Linder:2002et}) with $\Upsilon(a)=(1-\Omega_{m})a^{-3(1+w_o+w_a)}e^{3w_a(a-1)}$ (see, e.~g.~\cite{Scherrer:2015tra}). The latter  arising from the assumption of a time-dependent dark energy equation of state given by $w_\text{DE}(a)=w_o+w_a (1-a)$. As shown in \cite{Carroll:2003wy}, all $f(R)$ theories have a constant strength in the Yukawa potential, namely $2\beta^2=1/3$. Hence, the theory is characterized by two mass-related parameters: $f_{R_0}$ controls the reach of the associated fifth force, while the power index $n$ determines its modulation over time.

\begin{figure}[htbp]
    \centering
    \includegraphics[width=0.8\textwidth]{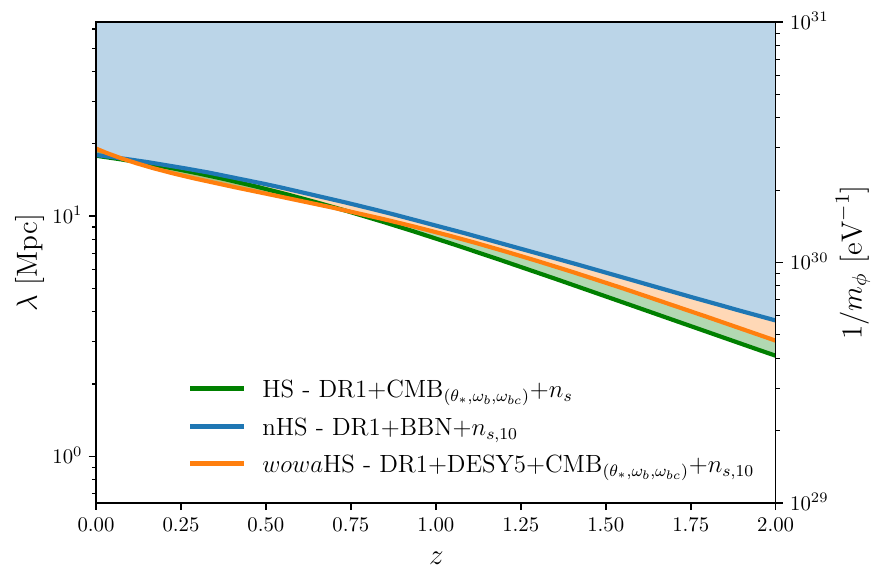}
    \caption{Constraints on the MG Yukawa scale as a function of redshift, using DESI DR1(FS+BAO), CMB compressed information, and DES-Y5 data. Our baseline model uses the map between the Yukawa potential and the unscreened $f(R)$ Hu-Sawiki model to constrain the fifth force interaction scale $\lambda$, or equivalently, $1/m_\phi$. Coloured lines represent the 95\% bound for the baseline model (HS, green), an extension with full HS theory (nHS, blue) and a dynamical equation of state for dark energy in the background ($wowa$HS, orange). The higher shaded region for each model depicts the ruled out values. 
    }
    \label{fig:m_phi}
\end{figure}

Our analysis is performed on power-spectra measurements of different DESI targets over the redshift range $0.1<z<2.1$. We contrast the monopole and quadrupole of these Fourier space correlations to the MG EFT model, finding consistent results among the different DESI targets and redshift bins. In summary, we find that the clustering data prefer GR. As a result, and using minimal BBN and CMB prior information for $\omega_b$ and $n_s$, we are able to constrain the mass scale $m_\phi$ at redshift zero to 
\begin{eqnarray}
    m_\phi &>& 3.60\times10^{-31} \text{ eV  (95\% c.l.)}\qquad \text{DESI DR1(FS)},\\
    m_\phi&>&3.57\times10^{-31} \text{ eV  (95\% c.l.)}\qquad \text{DESI DR1(FS+BAO)},
\end{eqnarray}
for FS only and combined with BAO, respectively.
When combined with compressed-CMB information, defined by additional Planck priors in $\theta_*,\omega_{b},\omega_{bc}$, one gets
\begin{equation}
    m_\phi> 3.59\times10^{-31} \text{ eV  (95\% c.l.)}\qquad \text{DESI DR1(FS+BAO)+CMB$_{(\theta_{*},\omega_{b},\omega_{bc})}$}, 
\end{equation}
while adding the DES-Y5 supernova sample leads to the tightest constraint 
\begin{equation}
    m_\phi> 3.36\times10^{-31} \text{ eV  (95\% c.l.)}\qquad \text{DR1(FS+BAO)+CMB$_{(\theta_{*},\omega_{b},\omega_{bc})}$+DES-Y5}.
\end{equation}
 All of these constraints translate into a length scale, $\lambda=m_\phi^{-1}$, which is smaller than about 18 Mpc. Additional combinations of datasets and modeling choices are summarized in \cref{fig:m_phi}, while details on the methodology, validation tests, and further results are provided in the following sections. Moreover, the redshift evolution of this mass scale defined by \cref{eq:HS-mass-relation} can also be appreciated in the figure for different modeling and dataset combinations.

The paper is organized as follows. \cref{sec:model} describes the details of the modeling employed for this Full-Shape analysis, including the mapping of the Yukawa fifth-force potential to the unscreened HS model. \cref{sec:data} provides a recount of the used datasets, including the DESI DR1 data as well as from non-DESI cosmological observations. \cref{sec:methodology} describes our methodology, defining the baseline model and extensions to it. Results are given in \cref{sec:results}, providing validations of the baseline model, and the impact of extensions to this baseline model. We further study the degeneracies between the MG parameter $f_{R_0}$ and the sum of neutrino masses in \cref{sec:MGmnu}. At last, in \cref{sec:conclusions}, we conclude with some Final thoughts on this work.

\section{Full-Shape modeling for scale-dependent MG}\label{sec:model}

In recent years, important progress has been made in describing matter density fluctuations in the mildly non-linear regime using an EFT approach \cite{Bernardeau:2001qr,McDonald:2009dh,Baumann:2010tm,Carrasco:2012cv,Pajer:2013jj,Assassi:2014fva,Mirbabayi:2014zca,Vlah:2015zda,Aviles:2018thp,Ivanov:2019pdj,DAmico:2019fhj,Chen:2020zjt,Aviles:2020cax,Blas:2015qsi}. This framework combines physically motivated IR resummations \cite{Senatore:2014via,Mirbabayi:2014gda,Baldauf:2015xfa,Ivanov:2018gjr} and UV counterterms \cite{Baumann:2010tm,Carrasco:2012cv} with a consistent bias model for the galaxy-dark matter connection (see e.g. \cite{McDonald:2009dh,Assassi:2014fva,Desjacques:2016bnm}). Together, these ingredients provide an accurate description of the auto-power spectra of DESI tracers up to scales of $k\sim 0.2 \hmpci$, as demonstrated in \cite{DESI2024.V.KP5} and the references therein. This EFT-based description of DESI data has been performed for $\Lambda$CDM and several extensions, including the $wowa$ parametrization for dynamical dark energy \cite{DESI2024.VII.KP7B} and a free mass parameter for three degenerate neutrino mass eigenstates \cite{Elbers:2025vlz}. While an analysis of scale-independent departures from GR using DESI data has been carried out within this framework \cite{KP7s1-MG}, the investigation of scale-dependent departures remains open; addressing this gap is the primary goal of the present work.

In the presence of a fifth-force, the scale dependence of the velocity divergence field, $\theta^{(1)}(\mathbf{k},t)$, is different from that of the density fluctuation, $\delta^{(1)}(\mathbf{k},t)$, at lowest order in perturbation theory, namely\footnote{We define $\theta \equiv -\nabla \cdot \mathbf{v}/(aHf_0)$, with $\mathbf{v}$ the peculiar velocity, $H$ the Hubble factor and $f_0$ the growth rate as $k \to 0$.}
\begin{equation}
    \theta^{(1)}(\mathbf{k},t)=\frac{f(k,t)}{f_0(t)}\delta^{(1)}(\mathbf{k},t),
\end{equation}
where the scale-dependent growth rate 
\begin{equation}
    f(k,t)=\frac{d\log D_+(k,t)}{d\log a(t)},
\end{equation}
and $f_0=f(k=0,t)$ is the linear growth rate at the largest scales. The linear growth functions $D_+$ encodes the fifth-force through the modified growth equation
\begin{align}\label{Dplus}
 \ddot{D}_+(\mathbf{k}, t) + 2 H \dot{D}_+(\mathbf{k}, t) - \frac{3}{2} H^2 \Omega_{m} (a) \left (1 + \frac{2\beta^2\,\mathbf{k}^2}{\mathbf{k}^2+m_\phi^2 a^2}\right) D_+(\mathbf{k}, t) = 0,
\end{align}
where the additional mass scale $m_\phi$ is, generically, time-dependent.  Notice that \cref{Dplus} reduces to the $\Lambda$CDM scenario in the limit $k\rightarrow 0$. In contrast, for $k>a m_\phi$ the additional scalar field mediated force enhances the strength of gravity by a factor $1+2 \beta^2$. As a result, the growth function and factor monotonically interpolate between these two regimes. 

For MG theories that only affect the late-time structure formation, as is generally the case of those used as alternatives to a cosmological constant, the linear power spectrum can be computed by the rescaling \cite{Aviles:2017aor}
\begin{equation} \label{PLMG}
 P_L(k,z) = \left( \frac{D_+(k,z)}{D_+^\text{$\Lambda$CDM}(z_0)} \right)^2 P_L^\text{$\Lambda$CDM}(k,z_0),   
\end{equation}
where $D_+^\text{$\Lambda$CDM}$ and $P_L^\text{$\Lambda$CDM}$ are the linear growth function and power spectrum computed in the $\Lambda$CDM model. \Cref{PLMG} represents a considerable computational simplification since one can use a Boltzmann code in $\Lambda$CDM to perform the early times linear evolution, and then rescale the transfer functions with the linear growth functions in MG. If this is not possible one should use a linear solver as, e.g., \textsc{ISiTGR} \cite{Garcia-Quintero:2019xal,Dossett_2011}.\footnote{\url{https://github.com/mishakb/ISiTGR}}

A first prescription to include scale-dependent MG in this EFT description is to track the $k$-dependence on the growth function and the growth rate into the linear and higher-order perturbative kernels. This is the \texttt{fkpt} approximation, developed in \cite{Aviles:2021que,Rodriguez-Meza:2023rga}, which is also applied to study the suppression of the power spectrum due to massive neutrinos \cite{Aviles:2021que,Noriega:2022nhf,Noriega:2024lzo}. 

We can write the power spectrum at 1-loop in PT as an expansion of the form \cite{Aviles:2020wme}
\begin{equation}
    P(k,\mu)=\sum_{m=0}^n\sum_{n=0}^4 \mu^{2n}f_0^m I_{mn}(k),
\end{equation}
where $\mu$ is the cosine angle between the wave vector and the line-of-sight, $f_0\equiv f(k=0,t)$, and the functions $I_{mn}(k)$ come in two kinds:
\begin{align}
      & I_{mn}(k) = \int \Dk{p}  \mathcal{K}(\vp,\vk-\vp) P_L(p) P_L(|\vk-\vp|), \qquad  \text{(type 1, e.g. $P_{22}$)}, \label{ImnP22}\\
      & I_{mn}(k) = P_L(k) \int \Dk{p}  \mathcal{K}(\vk,\vp)  P_L(\vp), \qquad \qquad \qquad  \text{(type 2, e.g. $P_{13}$)},\label{ImnP13}
\end{align}
where $\mathcal{K}$ denotes the \texttt{fkpt} kernels. These kernels are known analytically up to the functions $D_+(k,t)$ and $f(k,t)$, which depend on the cosmology. Explicit expressions are given in \cite{Rodriguez-Meza:2023rga,Noriega:2022nhf}. As is standard in PT, the integrands in \cref{ImnP22,ImnP13} are rotationally invariant, hence the integration domains reduce to two dimensions.

Counterterms of the form $k^2 P_L(k)$ are required to renormalize the UV behavior of the type 2 one-loop corrections. In this work we only use the monopole and the quadrupole of the power spectrum, hence the relevant counterterms are controlled by the nuisance parameters $\alpha_0$ and $\alpha_2$, corresponding to powers of $\mu^0$ and $\mu^2$, respectively. This yields
$P_\text{ctr}= (\alpha_0 +\alpha_2 \mu^2 ) k^2 P_L(k)$.
Furthermore, a shot noise contribution is added to the power spectrum as $P_\text{shot} = \bar{n}^{-1}\big( \alpha_0^\text{shot} + (k \mu)^2 \alpha^\text{shot}_2\big)$, where $\bar{n}$ is the mean galaxy number density. We also include IR resummations \cite{Senatore:2014via,Baldauf:2015xfa,Vlah:2015zda}, which accounts for the physical, non-perturbative effect of large-scale coherent motions, which smooth the BAO signal by displacing particles. Technically, this is implemented using the method of  \cite{Ivanov:2018gjr}, by decomposing the linear power spectrum into a wiggly part, $P_w$, and a smooth, no-wiggle part, $P_{nw}$ (i.e., $P=P_w+P_{nw}$), which couple differently to long-wavelength modes.

To complete the theory, we add a consistent bias model based on \cite{McDonald:2006mx,McDonald:2009dh}. 
In our case, the relevant bias parameters are the linear and second order local biases, $b_1$ and $b_2$, respectively, the tidal bias $b_s$, and the third order non-local bias $b_{3nl}$. Following the results of co-evolution theory \cite{Chan:2012jj,Baldauf:2012hs,Saito:2014qha}, we opt for fixing $b_{3nl}=32(b_1-1)/315$ always, while leave the option of fixing or not $b_s$ to its co-evolution value, $b_s=-4(b_1-1)/7$. While MG theories lack a complete bias basis \cite{Desjacques:2016bnm}, the inclusion of curvature-related terms has been shown to successfully describe tracer evolution in various MG models \cite{Aviles:2020wme,Rodriguez-Meza:2023rga}.

\subsubsection*{Map to $f(R)$ theories}

A second prescription in our scale-dependent MG analysis is to use the mapping of the $f(R)$ HS model to the Yukawa potential given by \cref{eq:HS-mass-relation}. The action for the HS model is given by $S=\frac{1}{2}M_{Pl}^2\int d^4x\sqrt{-g}(R+f(R))$, where the function $f(R)$ is defined as \cite{Hu:2007nk}
\begin{equation}
f(R)= -M^2 \frac{c_1 (R/M^2)^n}{c_2 (R/M^2)^n + 1},
\label{eq:hu-sawicki-model}
\end{equation}
with $c_1$ and $c_2$ dimensionless parameters. The energy scale is chosen to be $M^2 = H_0^2 \Omega_{m}$ to match the observed dark energy density. In this model, at large 4D curvature ($R \gg M^2$), $f(R)$ approaches a constant, recovering General Relativity (GR) with an effective cosmological constant. 
To mimic the background expansion history of $\Lambda$CDM, the parameters must satisfy the condition $c_1/c_2 = 6\Omega_\Lambda/\Omega_{m}$ \cite{Hu:2007nk}. We define the Ricci scalar in the background as $\bar{R}\equiv 6\dot{H}+12H^2$. For $\bar{R} \gtrsim 40M^2$, which is valid throughout the cosmic history relevant for DESI, the model simplifies and the scalar degree of freedom, $f_R \equiv df(R)/dR$, can be expressed as
\begin{equation} \label{eq:fR_approx}
f_R \simeq f_{R_0} \left( \frac{\bar{R}_0}{\bar{R}(a)} \right)^{n+1},
\end{equation}
where $f_{R_0}$ is the value of the field today.

The dynamics of this scalar field are governed by the trace of the modified Einstein equations. In the quasi-static limit, and after subtracting the background contribution, we obtain the Klein-Gordon equation for its perturbation, $\delta f_R$ \cite{Li:2017xdi}:
\begin{equation}\label{eq:kg-fR}
\frac{3}{a^2} \nabla^2 \delta f_R = - 8 \pi G \bar{\rho}_m \delta_m + \delta R,
\end{equation}
where $\delta_m$ is the matter overdensity and $\delta R= R - \bar{R}$. By expanding the potential $\delta R$ as a Taylor series in $\delta f_R$, $\delta R = \sum_j \frac{1}{j!} M_j (\delta f_R)^j$ \cite{Koyama:2009me}, the mass of the field can be identified as $m_\phi = \sqrt{M_1/3}$. For the HS model with $n=1$,  $M_1$ is given by
\begin{equation} \label{eq:M1}
M_1(a) = \frac{3}{2} \frac{H_0^2}{|f_{R_0}|} \frac{(\Omega_{m} a^{-3} + 4 \Omega_\Lambda)^3}{(\Omega_{m} + 4 \Omega_\Lambda)^2} = 3 m^2_\phi.
\end{equation}
As a reminder, the lenght scale at which GR corrections become important, $\lambda$, is the inverse of scalar field mass, namely $\lambda=1/m_\phi$. The solution to the perturbed field equations in Fourier space then leads to a modification of the Poisson equation, which in the linear regime takes the form of a Yukawa potential with a coupling of $1/3$ and a comoving interaction range given by $k_\mathrm{MG}^{-1} = (a m_\phi)^{-1} = a^{-1}\sqrt{3/M_1}$.


At this level, the use of this mapping represents a useful parameterization to place constraints on the MG scale $k_\mathrm{MG}$, but does not restrict the analysis to that particular $f(R)$ model. The screening mechanism which restores GR on small scales is inherent to this class of models and arises from the higher-order terms in the Taylor expansion of the potential, $\delta R = \sum_{j>0}\frac{1}{j!} M_j (\delta f_R)^j$. By truncating this expansion at linear order in $\delta f_R$, we obtain the Yukawa description in \cref{eq:HS-mass-relation}, which captures the linearized fifth force but neglects the non-linear effects responsible for screening. For our FS analysis, which focuses on linear to mildly non-linear scales, a conservative maximum scale cut ensures that we remain in the regime where these non-linear corrections are subdominant and can be accounted for the EFT counterterms (as shown in \cite{Rodriguez-Meza:2023rga}), such that the linear Yukawa description remains valid.

\begin{figure}
    \centering
    \includegraphics[width=5 in]{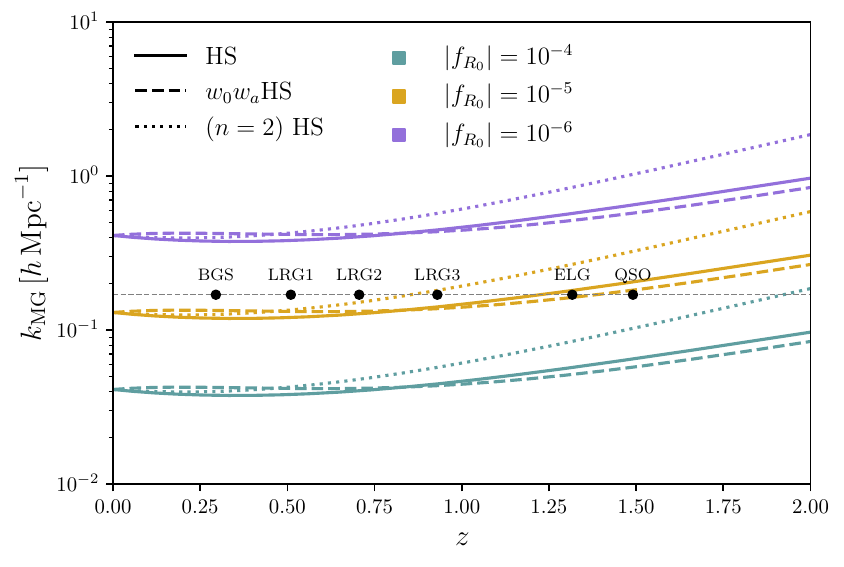}
    \caption{Modified gravity scale $k_\text{MG}=a m_{\phi}(a)$ as a function of redshift, computed using \cref{eq:HS-mass-relation} and the cosmological best-fit parameters from \cite{DESI.DR2.BAO.cosmo}. Solid lines show the minimum MG scale for the values $f_{R_0}=10^{-4}, 10^{-5}$ and $10^{-6}$, assuming $n=1$, while dotted lines assume $n=2$. We also consider a $wowa$-background evolution, assuming also $n=1$, which we show with dashed lines.  The dashed gray horizontal line indicates the conservative maximum scale used in the FS analysis, $k=0.17 \hmpci$, such that only models whose curves lie below this line are accessible to the galaxy power spectrum. Dots mark the effective redshift bins of DESI for the different tracers.}
    \label{fig:mg_scale}
\end{figure}

To gain insight into the scale at which modifications to GR become relevant for DESI, we explore several scenarios for the power index $n$. In our baseline analysis, we fix $n=1$, the most common choice in the literature, and constrain the allowed values of $f_{R_0}$, while remaining on scales where screening effects are degenerate with EFT counterterms. The corresponding MG scale, $k_\text{MG}=a m_{\phi}$, is shown in \cref{fig:mg_scale} for different values of $f_{R_0}$, across DESI's redshift bins. The horizontal dashed gray line indicates the conservative maximum wavenumber we use,  $k_\text{max}=0.17 \hmpci$ used in the baseline analysis, such that only models whose curves lie below this line are accessible to the galaxy power spectrum. To assess the robustness of our constraints to the assumed functional form, we also consider two additional cases: (i) fixing $n=2$, and (ii) allowing $n$ to vary freely. In the figure, we also show curves for $n=2$ and different values of $f_{R_0}$. For $n=1$, we use $\Lambda$CDM and $wowa$ models to describe the background evolution. A general $wowa$ background is incompatible with the Hu–Sawicki model, but is accommodated within the broader Horndeski framework, which is sufficient to probe the scale-dependent modification of gravity we pursue. Notably, the redshift range $0<z<1$ provides the strongest constraining power for scale-dependent modifications of GR with DESI.

There is a vast literature on testing $f(R)$ models in cosmology (see, for example, the reviews of \cite{Sotiriou_2010,De_Felice_2010,Koyama_2016,Ishak_2018}). In this context, the constraints in $f_{R_0}$ obtained in this work should be contrasted with
those obtained using other methods in the literature, often assuming the value of $n$ to be one. At present, the strongest upper limit in the HS model arises from astrophysical scales, $|f_{R_0}| < 10^{-8}$ (95\% confidence) \cite{Landim:2024wzi}, which is significantly tighter than constraints from cosmological observations. On the cosmological side, the most restrictive bound comes from a recent joint analysis of SPT clusters and Planck 2018 data, yielding $|f_{R_0}| < 10^{-5.32} = 4.79 \times 10^{-6}$ \cite{SPT:2024adw}. A more recent cluster abundance study using eROSITA alone finds a weaker constraint, $|f_{R_0}| < 10^{-4.12} = 7.59 \times 10^{-5}$ \cite{Artis:2024eco}. From large-scale structure probes, the tightest limit from weak-lensing peak statistics is $|f_{R_0}| < 10^{-4.82} = 1.51 \times 10^{-5}$ from CFHTLenS \cite{Liu:2016xes}; the authors note that including CMB priors on $\Omega_m$ and $A_s$ tightens the bound to values below $10^{-5}$. A FS analysis from BOSS DR12 gives $f_{R_0} < 10^{-4.82} = 1.53 \times 10^{-5}$ \cite{Aviles:2024zlw}, which is somewhat stronger than our result, although the analysis assumes co-evolution theory for bias, tighter priors on EFT parameters, and fixes $n_s$ to the Planck value. Finally, a joint cosmic shear analysis combining DES-Y3, KiDS-1000, and HSC-Y3, together with CMB and BAO data \cite{Bai:2024hpw}, reports $|f_{R_0}| < 10^{-4.98} = 1.05 \times 10^{-5}$.

\section{Datasets} \label{sec:data}

The main observational data in our FS analysis is the DESI DR1 \cite{DESI2024.I.DR1}, available at \url{https://data.desi.lbl.gov/doc/releases/dr1/}. The DESI wide-angle instrument is capable of taking around 5000 simultaneous spectra using robotic-fiber positioners over the optically corrected focal plane \cite{FiberSystem.Poppett.2024, Corrector.Miller.2023,FocalPlane.Silber.2023}, and ten high-resolution spectrographs \cite{Spectro.Pipeline.Guy.2023}. DR1 comprises a survey campaign of 13 months after the main survey started in May 2021 \cite{SurveyOps.Schlafly.2023}, with around 14 million galaxies and quasars, following a target selection algorithm \cite{TS.Pipeline.Myers.2023} using the Legacy Survey photometric catalogs \cite{LS.Overview.Dey.2019,LS.dr9.Schegel.2024} and a survey validation campaign \cite{DESI2023a.KP1.SV}. 

\subsection{DESI DR1}\label{sec:DESI.DR1}

For the FS analysis, we use the monopole and quadrupole moments of the power spectra of BGS \cite{BGS.TS.Hahn.2023}, LRG \cite{LRG.TS.Zhou.2023}, ELG \cite{ELG.TS.Raichoor.2023} and QSO \cite{QSO.TS.Chaussidon.2023} targets, measured over the following z-bins: BGS - $0.1 < z < 0.4$, LRG1 - $0.4 < z < 0.6$, LRG2 - $0.6 < z < 0.8$, LRG3 - $0.8 < z < 1.1$, ELG2 - $1.1 < z < 1.6$, and QSO - $0.8 < z < 2.1$. A description of the FS analysis in $\Lambda$CDM for these targets, including information on random catalogs, window functions, weights and blinding scheme can be found in \cite{DESI2024.V.KP5} (with basic properties shown in its Table 1), together with its corresponding cosmological analysis \cite{DESI2024.V.KP5,DESI2024.VII.KP7B}. Covariance matrices were obtained from 1000 EZmocks realizations, with corrections based on higher resolution simulations from the Abacus suite and analytical methods using \textsc{RascalC} \cite{2023MNRAS.524.3894R, KP4s7-Rashkovetskyi, KP4s6-Forero-Sanchez}. 

Of particular relevance for this work is the consistency of different EFT approaches for DESI's Full-Shape standard analysis \cite{KP5s3-Noriega, KP5s4-Lai, KP5s5-Ramirez, KP5s1-Maus, KP5s2-Maus}. In the Bayesian framework, this EFT modeling introduces nuisance parameters, capturing the bias, counterterms and shot noise expansions, which should be marginalized over to derive the cosmological parameters. These nuisances are often  
degenerate with the cosmological parameters when weakly constrained, leading to non-physical shifts in the inferred cosmology which strongly departure from the Maximum Posterior Value (MAP) \cite{Simon:2022lde, Gomez-Valent:2022hkb, Carrilho:2022mon, Hadzhiyska:2023wae, Holm:2023laa, KP5s2-Maus,Tsedrik:2025hmj}. These projection effects are particularly relevant to extended cosmologies, such as the $wowa$ parametrization \cite{KP5s2-Maus, DESI2024.V.KP5}. There are a few strategies to mitigate these effects, including simulation based priors \cite{DESI2024.V.KP5}, HOD-informed priors \cite{DESI:2025wzd, Ivanov:2024xgb},  re-parametrizations that break the degeneracies between nuisances and the cosmological parameters \cite{KP5s2-Maus,Tsedrik:2025hmj,Paradiso:2024yqh}, among others. As a result, it is important to assess whether new variables in extended cosmological models lead to additional projection effects.

We further combine the DESI FS analysis with DESI BAO data from DR1, which in turn is based on post-reconstruction measurements in configuration space of the clustering in the discrete DESI targets described above. We use anisotropic BAO fits for the LRG and ELG samples, while isotropic fits for the BGS and QSO targets. Details of the BAO measurements, blinding, validation and cosmological interpretations can be found in \cite{DESI2024.II.KP3, DESI2024.III.KP4, DESI2024.VI.KP7A} and references therein.

\subsection{External (non-DESI) cosmological data}\label{sec:CMBc}
A minimal input for a DESI stand alone analysis is needed, which can be taken from different sources. We use information in the baryonic density, $\omega_b$, from Big Bang Nucleosynthesis (BBN), and a conservative prior in the spectral index of primordial fluctuations, $n_s$, from the CMB. We give details of these two particular choices in the following sections.

Furthermore, to capture a relevant sector of CMB data for our analysis, we follow the same idea as in the DESI BAO DR2 analysis \cite{DESI.DR2.BAO.cosmo} and use CMB-compressed (geometric) information, defined by correlated Gaussian priors in the angular distance to recombination $\theta_*$, and densities $\omega_b$ and $\omega_{bc}=\omega_b+\omega_c$. These are obtained after marginalizing over the Integrated Sachs–Wolfe effect, lensing contributions, and other late-time effects, which, based on \cite{Lemos_2023}, result in the Gaussian priors with mean
\begin{equation}\label{eq:cmb_priors}
    \mathbf{\mu}(\theta_*,\omega_{b},\omega_{bc})=\begin{pmatrix}
        0.01041\\ 0.02223\\0.14208
    \end{pmatrix},
\end{equation}
and covariance
\begin{equation}\label{eq:cmb_covs}
    \mathbf{C}=10^{-9}\begin{pmatrix}
        0.006621 & 0.12444 & -1.1929\\
        0.12444 & 21.344 & -94.001\\
        -1.1929 & -94.001 & 1488.4
    \end{pmatrix}.
\end{equation}
This information is simply labeled as CMB$_{(\theta_*,\omega_{b},\omega_{bc})}$ throughout this work. Moreover, to capture the impact of a CMB likelihood with these compressed information, we choose Planck's prior on $n_s$ instead of the 10 times wider prior used in DESI alone models.\footnote{There is also an alternative consistent way of defining CMB compressed information based on $\omega_b$, and CMB shift parameter $R$ and angular location $l_A$ \cite{bansal2025expansionhistorypreferencesdesidr2}.} As previously explained, the CMB information is not that sensitive to this MG scale, thus the validation of the compressed CMB likelihood is the same as for the $\Lambda$CDM case, which is discussed in Appendix A of \cite{DESI.DR2.BAO.cosmo} in the context of the DESI analyses. In particular, in the class of models considered here, the modification mainly affects the late-time galaxy power spectra by enhancing them above the characteristic scale $k_\text{MG}$. Therefore, moving from compressed CMB information to the full CMB likelihood is not expected to significantly change the constraints, except possibly through a modest additional sensitivity from CMB lensing, which probes the late-time growth of structure. 


In addition, we combine DESI and compressed CMB with the Year 5 likelihood from the Dark Energy Survey collaboration \cite{descollaboration2025darkenergysurveycosmology} (referred as DES-Y5 from now onward). This likelihood was used in the DESI DR1 analysis \cite{DESI2024.VII.KP7B} and contains 1635 SNIa light-curves in the redshift range $0.1<z<1.3$. 
Although more recent calibrations and analyses of DES-Y5 lead to weaker tensions with $\Lambda$CDM when combined with DESI and CMB data (see, for example, the analysis of \cite{DES:2025sig}), we use the DES-Y5 likelihood as an illustrative comparison case rather than as a preferred baseline. Our aim is to assess the robustness of the inferred MG constraints to changes in the assumed background expansion history, and DES-Y5 provides a useful example in that respect. For completeness, we also explore the impact of using DES-Dovekie \cite{DES:2025sig}, Pantheon+ \cite{Scolnic_2022} and Union3 \cite{rubin2025unionunitycosmology2000} supernova datasets.

\section{Methodology} \label{sec:methodology}

To constrain the scale-dependent MG model, we perform a Bayesian parameter inference for a baseline model with additional extensions. We use a Gaussian likelihood and the Metropolis–Hastings MCMC sampler (\cite{Lewis_2013,PhysRevD.66.103511}) implemented in \textsc{desilike}\footnote{\url{https://github.com/cosmodesi/desilike/}}
 or \textsc{cobaya}\footnote{\url{https://cobaya.readthedocs.io/}}
 \cite{Torrado_2021,2019ascl.soft10019T}.
In our models, the early Universe is very close to $\Lambda$CDM; hence, we use the Boltzmann code \textsc{CAMB} \cite{Lewis_2000} to obtain the linear power spectrum, which is then rescaled using \cref{PLMG}. For most models, we use the \textsc{desilike} pipeline, which relies on an emulator to interpolate between model parameters. We validate the emulator by running chains without it. Only when including the DES-Y5, or other SNe datasets, we employ \textsc{cobaya}. We use the \textsc{GetDist}\footnote{\url{https://getdist.readthedocs.io/en/latest/}.} code \cite{Lewis_2025} to analyse the chains, produce plots and derive the final constraints. All posterior contours correspond to the 68\% ($\simeq 1\sigma$) and 95\% ($\simeq 2\sigma$) credibility intervals.

\subsection{Baseline and extension models}
We define the baseline model as the FS analysis of DESI data only using the HS parametrization discussed in \cref{sec:model} with $n=1$ fixed. This baseline model is naturally compared with $\Lambda$CDM and considering the same assumptions for the non-MG features. 
The FS analysis is restricted to the monopole and quadrupole of the power spectrum over the interval $0.02<k/(\hmpci)<0.17$. We adopt the conservative choice $k_{\rm max}=0.17\,h\,{\rm Mpc}^{-1}$ because, as shown in Ref.~\cite{Rodriguez-Meza:2023rga}, over this range the screening contribution is largely degenerate with the EFT counterterms entering the FS model, and can therefore be neglected in our analysis.
The three neutrino mass-states are degenerated and fixed to 0.06 eV, which is consistent with neutrino oscillation experiments (see, for example, \cite{Esteban:2024eli}). We also explore the scenario where the sum of the neutrino masses is a free parameter. Additionally, we opt for a 10-fold wider Gaussian prior in $n_s$ from the Planck 2018 result \cite{Planck:2018vyg}, which we refer as $n_{s,10}$ in plots. Finally, the baseline model also includes a Gaussian BBN prior in $\omega_b$ from \cite{Cooke:2017cwo}. 

In order to validate the baseline model, we also consider modest variations. We allow the $n$ parameter in the HS model to be free and assess how accurate it is to fix it to one in the baseline model. Moreover, to estimate the impact of $k_\text{max}=0.17 \hmpci$ instead of the preferred value of $0.2\hmpci$ in DESI official analyses \cite{DESI2024.V.KP5,DESI2024.VII.KP7B}, we also run the baseline model up to $k_\text{max}=0.2 \hmpci$. Therefore, all FS analyses are run up to $k_\text{max}=0.17 \hmpci$ unless otherwise stated.

As previously noted, the modified gravity scale is primarily constrained by the Full-Shape analysis of the galaxy power spectrum. However, incorporating cosmological information from other observables can break degeneracies between model parameters, thereby potentially tightening the bounds on the MG mass $m_\phi$. To this extent, we add DESI BAO measurements for the same DR1 and the same targets and redshift bins. The model is compared to the FS and BAO measurements jointly, using the auto and cross covariances described in \cref{sec:DESI.DR1}. At the time of writing we lack cross covariances between the most recent BAO measurements from DR2 and FS from DR1, hence leaving that combination to the upcoming FS analysis with DR2.

Finally, we also consider extensions that include the CMB compressed likelihood defined in \cref{sec:CMBc}, and the DES-Y5 and DES-Dovekie supernova likelihoods. 

\subsubsection*{Priors}

\Cref{tab:priors} summarizes our choice of priors, including the nuisance parameters arising from the EFT and bias prescriptions.
For nuisances, we use the reparametrization of \cite{KP5s2-Maus},\footnote{This reparametrization was recently expanded to account for the change in amplitude of the power spectrum due to Alcock-Paczynski effect \cite{Tsedrik:2025hmj}.} which consist on vary and impose priors over the biases
\begin{equation}
 \tilde{b}_1 = b_1 \sigma_8, \quad 
  \tilde{b}_2 = b_2 \sigma_8^2 \quad \text{and} \quad   \tilde{b}_{s^2} = b_{s^2} \sigma_8^2.
\end{equation}
We further reparametrize the counterterms as 
\begin{equation} 
P_{ctr}(k,\mu) = (b_1 + f\mu^2)(b_1\tilde{\alpha}_0 + f\tilde{\alpha}_2\mu^2)k^2P_{L}(k),
\end{equation}
and impose priors on $\tilde{\alpha}_{0,2} \times \sigma_8^2$.
It is important to stress that these nuisance parameters, listed as one block in the table, are varied independently for each of the six redshift bins. Moreover, parameters that are linear in the power spectrum are analytically marginalized \cite{2010MNRAS.408..865T}. These are the counterterms $\tilde{\alpha}_{0,2}$ and stochastic parameters $\tilde{\alpha}^{\rm shot}_{0,2}$.

For $|f_{R_0}|$, we assume a flat prior in the interval $[0,10^{-4}]$. Although one could alternatively adopt a flat prior on $\log |f_{R_0}|$; however, Ref.~\cite{Rodriguez-Meza:2023rga} showed that the logarithmic choice can shift the recovered parameter toward lower values. In the same work, it was also shown that enlarging the prior range does not lead to appreciable changes in the results. Consistently, Ref.~\cite{Aviles:2024zlw} found that even for BOSS data, the inferred values of $|f_{R_0}|$ do not exceed $10^{-4}$.

\begin{table}[htbp] 
    \centering
     \renewcommand{\arraystretch}{1.1} 
    \begin{tabular}{|lcccl|}
    \hline
    Data or model & Parameter & Default & Prior & Comment\\  
    \hline 
    \multicolumn{5}{|l}{\textbf{\lcdm\ and MG shared parameters}}\\
    \rowcolor[HTML]{E5E5E5} & $H_{0} \; [\frac{\text{Km}}{\text{s}\,\text{Mpc}}]$ &---& $\mathcal{U}[20, 100]$ &---  \\
    \rowcolor[HTML]{E5E5E5} & $\ocdm$ &---& $\mathcal{U}[0.001, 0.99]$ &---\\
    \rowcolor[HTML]{E5E5E5} & $\ln(10^{10} A_{s})$ &---& $\mathcal{U}[1.61, 3.91]$ &---\\

    \hdashline
    \rowcolor[HTML]{E5E5E5} External priors & & && \\    
    
    \rowcolor[HTML]{E5E5E5} equal-mass states & $\sum M_\nu$ [eV] & 0.06  & ---  &  \\    
    \rowcolor[HTML]{E5E5E5} BBN & $\ob$ &---& $\mathcal{N}(0.02218, 0.00055^2)$ &  From \cite{Cooke:2017cwo}\\
    \rowcolor[HTML]{E5E5E5} $n_{s,10}$ & $n_{s}$ &---& $\mathcal{N}(0.9649, 0.042^2)$ & 10xPlanck \cite{Planck:2018vyg} \\       
    \hline     

    \multicolumn{5}{|l}{\textbf{MG parameters (Hu-Sawicki parametrization)}}  \\
    \rowcolor[HTML]{E5E5E5} & $f_{R_0}$ & $0$ & $\mathcal{U}[0, 10^{-4}]$ &---\\
    \rowcolor[HTML]{E5E5E5} & $n$ & $1$ & --- & fixed \\
    \hline

    \multicolumn{5}{|l}{\textbf{Nuisance parameters (EFT, bias model)}}\\ 
    \rowcolor[HTML]{E5E5E5}& $b_1 \sigma_8$ & --- & $\mathcal{U}[0, 3]$ & each $z$-bin\\    
    \rowcolor[HTML]{E5E5E5} & $b_2 \sigma_8^2$ & --- & $\mathcal{N}[0, 5^2]$ & each $z$-bin\\    
    \rowcolor[HTML]{E5E5E5} & $b_{s^2} \sigma_8^2$ & --- & $\mathcal{N}[0, 5^2]$ & each $z$-bin\\ 
    \rowcolor[HTML]{E5E5E5} & $b_{3nl}$ & $\frac{32}{315}(b_1-1)$ &  & co-evolution\\
    \rowcolor[HTML]{E5E5E5} & $\tilde{\alpha}_0 \sigma_8^2$ & --- & $\mathcal{N}[0, 12.5^2]$ & analytic\\
    \rowcolor[HTML]{E5E5E5} & $\tilde{\alpha}_2 \sigma_8^2$ & --- & $\mathcal{N}[0, 12.5^2]$ & analytic\\
    \rowcolor[HTML]{E5E5E5} & $\alpha^{\rm shot}_0$ & --- & $\mathcal{N}[0, 2^2]$ & analytic\\
    \rowcolor[HTML]{E5E5E5} & $\alpha^{\rm shot}_2$ & --- & $\mathcal{N}[0, 5^2]$ & analytic\\

    \hline

    \textbf{Extensions} 
    & & & & \\

    Minimal freedom (m.f.) & $b_s$ & $-\frac{4}{7}(b_1-1)$ &  & co-evolution\\
        
    $(n_{s})$ & $n_{s}$ &---& $\mathcal{N}(0.9649, 0.0042^2)$ & Planck \cite{Planck:2018vyg} \\       
    
    n-free & $n$ & --- & $\mathcal{U}[0,5]$ & \\

    
    $wowa$ & $w_0$ & --- & $\mathcal{U}[-3, 1]$ &---\\
     & $w_{a}$ & --- & $\mathcal{U}[-3, 2]$ &---\\

    CMB compressed & $w_b$ & --- & \multirow{3}{*}{$e^{-\frac{1}{2}\mu^T \mathbf{C}\mu}$} & $\mu$ and $\mathbf{C}$ are \\
      & $w_{cb}$ & --- & & given by Eqs.\\
    & $\theta_*$ & --- & & (\ref{eq:cmb_priors}) and (\ref{eq:cmb_covs}) \\

    \hline
    
    \end{tabular}
    \caption{Parameters and priors. Here, $\mathcal{U}$ denotes a uniform prior within the given range, while $\mathcal{N}(x, \sigma^2)$ represents a Gaussian distribution with mean $x$ and standard deviation $\sigma$. The default values recover the standard \lcdm\, model.  
    Our baseline settings, represented by the parameters shaded in gray, are consistent across all explored scenarios. Default entries are either fixed values by the chosen model or if unspecified in the code. CMB compressed priors are used with $n_s$, instead of $n_{s,10}$, to be consistent with a full CMB likelihood.}
    \label{tab:priors}
\end{table}

\section{Results on scale-dependent modified gravity constraints} \label{sec:results}

\begin{figure}[htbp]
 	\begin{center}
 	\includegraphics[width=5 in]{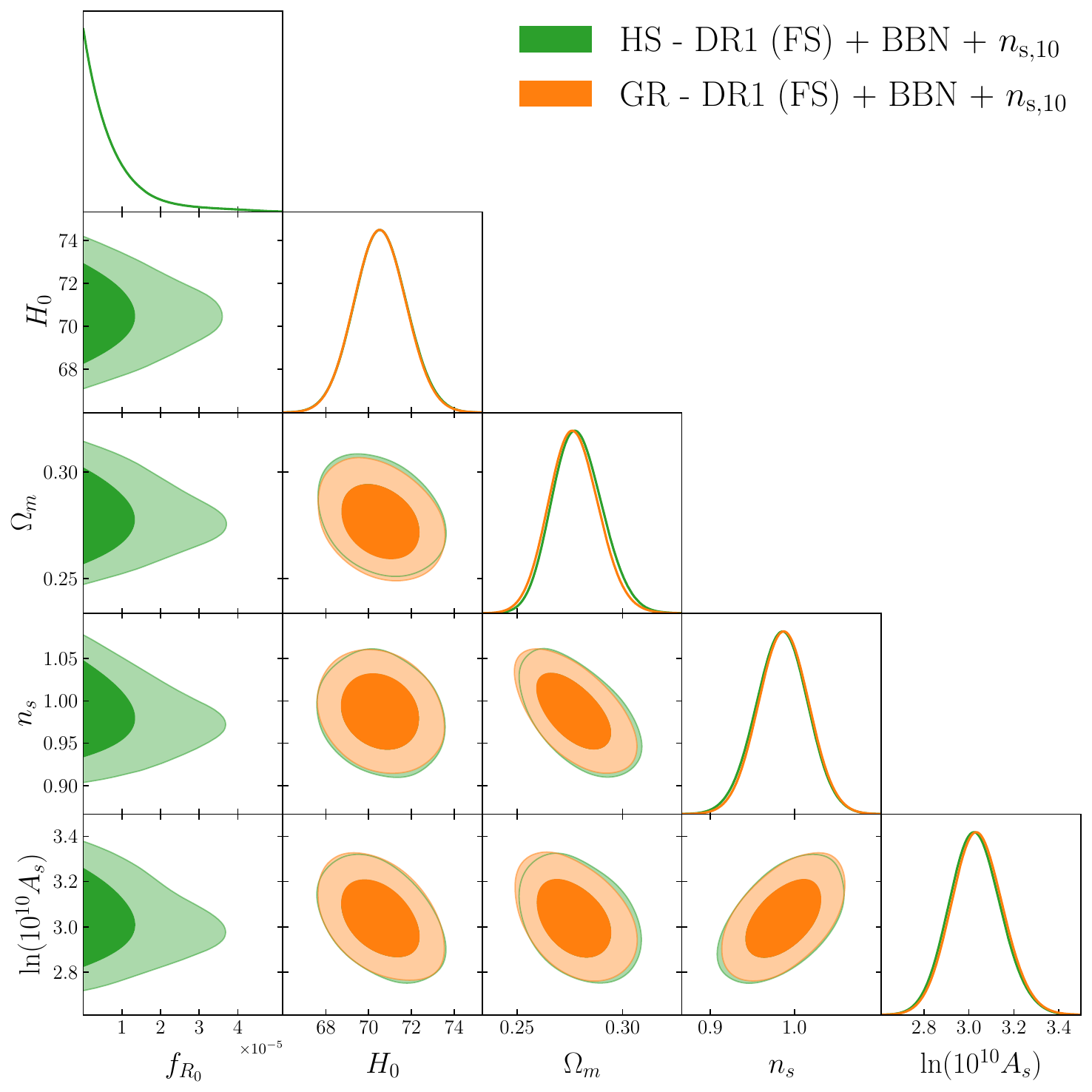}
 	\caption{Constraints on the GR and HS baseline models in the subspace spanned by the parameters $f_{R_0}$, $H_0$, $\Omega_m$, $n_s$ and $\ln(10^{10}A_{\mathrm{s}})$, assuming a $\Lambda$CDM background. The agreement of cosmological parameters between HS and GR to 2$\sigma$  is evidence of the modified gravity not introducing new projection effects and their orthogonality with $f_{R_0}$.} 
  \label{figure:triangular_desiy1_fs}
 	\end{center}
 \end{figure}

Our findings for the MG baseline model are summarized in \cref{figure:triangular_desiy1_fs}. We do not see any deviation from GR since the preferred value for the additional degree of freedom is $f_{R_0} <2.60\times 10^{-5}$ (95\% c.l.), and peaking at zero. As also appreciated in the figure, this additional parameter $f_{R_0}$ is practically orthogonal to the cosmological parameters, thus the cosmology is not affected with respect to the GR baseline model.  

\begin{figure}[htbp]
 	\begin{center}
 	\includegraphics[width=0.8 \textwidth]{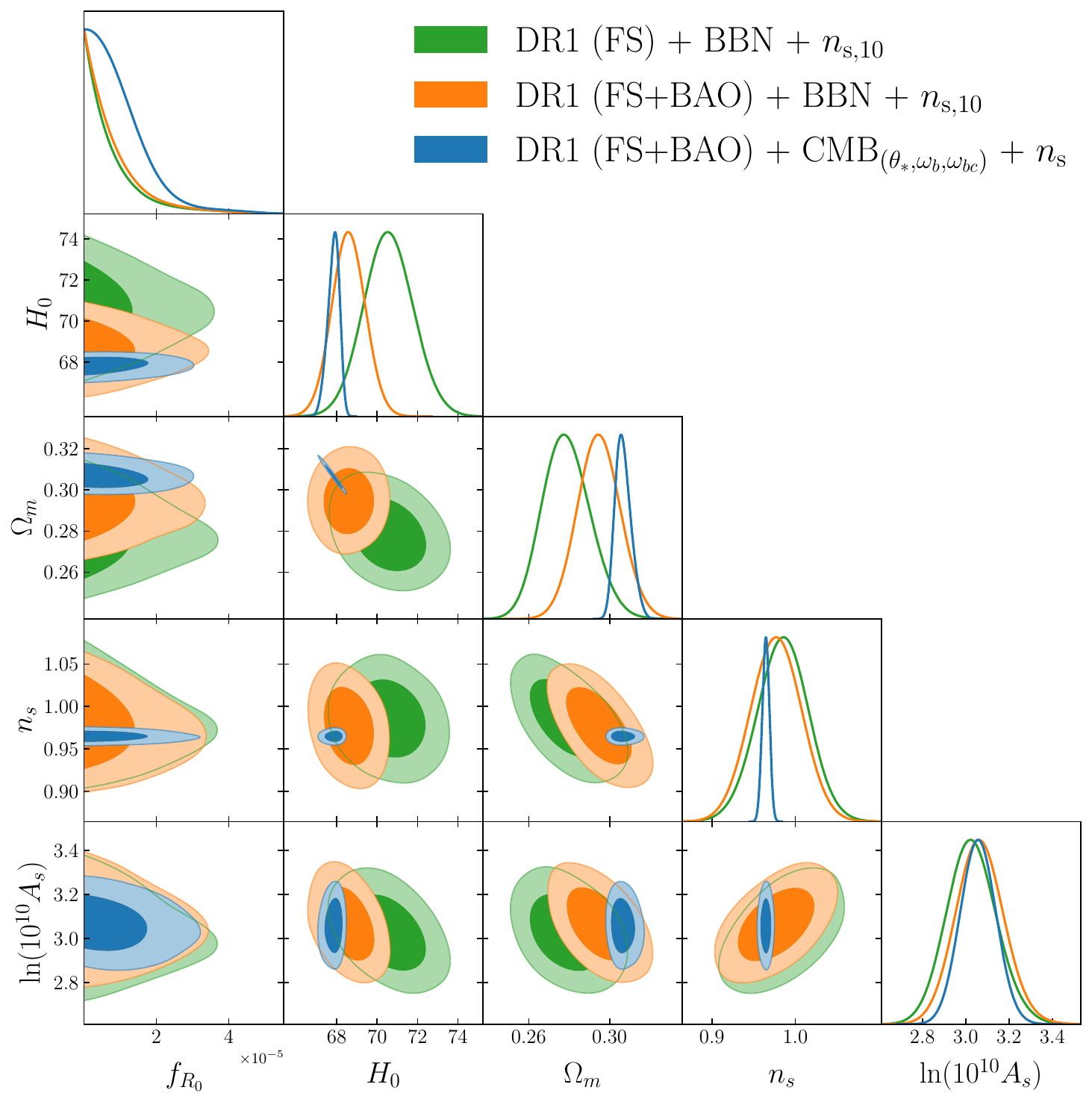}
 	\caption{
    2D posteriors on the HS $f(R)$ model assuming a $\Lambda$CDM background. 
    The green contours correspond to the baseline model, that uses the DR1 FS data up to a scale of $k_{\mathrm{max}}=0.17 \hmpci$. 
    Blue contours adds the DR1 BAO information, 
    improving the constraining power on the 
    cosmological parameters with the exception of $f_{R_0}$. Finally, the orange contours includes 
    $\mathrm{CMB_{(\theta_*,\omega_b,\omega_{bc})}}$.  
    Constraints on $f_{R_0}$ are robust between the different datasets. 
    }

  \label{figure:triangular_desiy1_fs_others}
 	\end{center}
 \end{figure}

Including BAO and performing a joint analysis with FS measurements does not alter the conclusions. The same happens when using the prior on $n_s$ reported by Planck, instead of the ten-folded choice of our baseline. At the level of $\chi^2$ per degree of freedom, all fits of the baseline with and without BAO perform similarly. The impact on these extensions to the baseline model are shown in \cref{figure:triangular_desiy1_fs_others}.

Furthermore, when combined with the CMB information, throughout CMB-compressed priors, the results do change slightly due to the well-known tension between DESI and the early universe (see, e.g., \cite{DESI.DR2.DR2}). Furthermore, the value of $\chi^2$ over the number of degrees of freedom ($\chi^2/\text{dof}$) is very high, indicating a poor fit between a theoretical model and the observed data, and the 1$\sigma$ region of the 1D projected posterior for $f_{R_0}$ widens, as observed in \cref{figure:triangular_desiy1_fs}. However, the overall 2$\sigma$ constraint on $f_{R_0}$ changes by no more than 4\%, confirming that it is the FS analysis that is most sensitive to scale-dependent departures of GR. Actually, the small gain in constraining $f_{R_0}$ when including different observables arise from the larger impact on the cosmological parameters. 

\begin{figure}[htbp]
    \centering
    \includegraphics[width=0.49\linewidth]{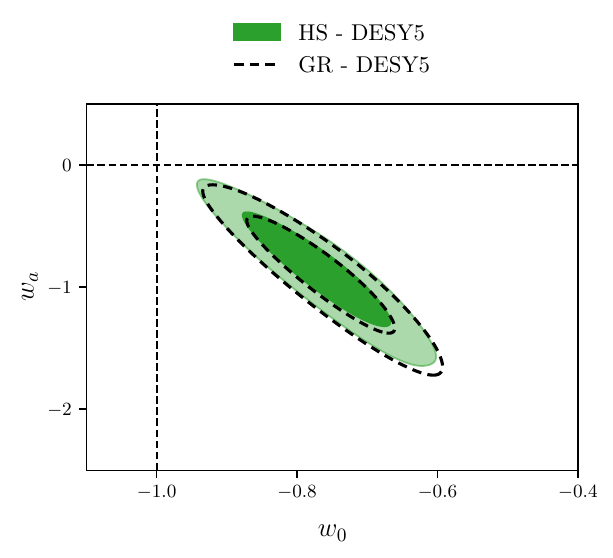}
    \includegraphics[width=0.49\linewidth]{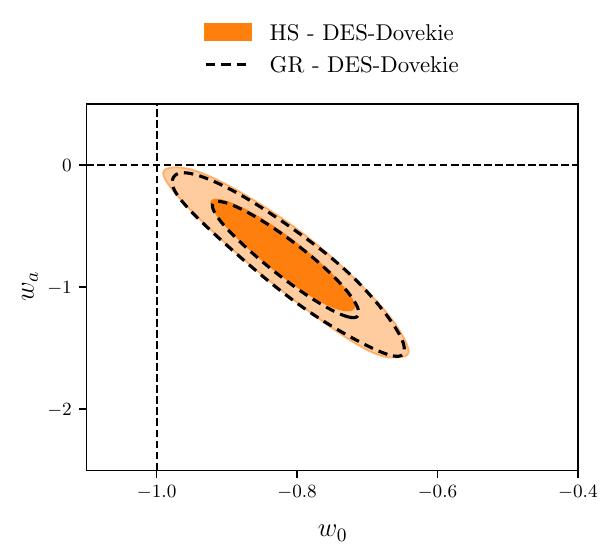}
   \hfill
    \caption{Constraints in the ($w_0,w_a$) plane using DES-Y5 and DES-Dovekie type Ia supernova likelihoods. Introducing the additional modified gravity parameter, $f_{R_0}$, does not alter the posterior distributions found in general relativity.}
    \label{fig:w_0w_a}
\end{figure}

Changing the background from $\Lambda$CDM to $wowa$ also does not play an important role, since one recovers almost the same cosmological parameters with or without MG. Actually, the choice of a dynamical dark energy in background relaxes the tension between DESI and CMB, returning the 1$\sigma$ region in the projected $f_{R_0}$ posterior to that found for the baseline model. By changing the background expansion to the $wowa$ parameterization, we are able to include SNIa data to the analysis. In the case of using all the datasets in the background $wowa$, including different 
SNIa datasets, one recovers the same tension with $\Lambda$CDM found in the DR1 analysis \cite{DESI2024.VII.KP7B}. \Cref{fig:w_0w_a} exemplifies this behaviour with DES-Y5 dataset, or with the more recent DES-Dovekie likelihood. Similar results are found for Pantheon+ and Union3 supernova datasets. 

\Cref{tab:results} summarises the bounds on the MG parameter $f_{R_0}$, the cosmological parameters $H_0$, $\ln(10^{10}A_s)$ and $\Omega_m$, as well as the $\chi^2/\text{dof}$ values of the fits.

\begin{table*}
\centering
\renewcommand{\arraystretch}{1.1} 
\resizebox{\linewidth}{!}{
    \begin{tabular}{|lccccc|}
    \hline
    Model and datasets & $f_{R_0}$ (95\% c.i.)& $\Omega_\mathrm{m}$ & $H_0$ & $\ln(10^{10}A_{\mathrm{s}})$ & $\chi^2$/dof\\
    \hline
    \textbf{GR - $\Lambda$CDM } &  &  &  & &\\
    \rowcolor[HTML]{E5E5E5} FS & - & $0.2769_{-0.0121}^{+0.0112}$ & $70.54\pm1.18$  & $3.038\pm0.111$  & 0.86 \\
    \hline
    \textbf{HS - $\Lambda$CDM} &  &  &  & & \\
    \rowcolor[HTML]{E5E5E5} FS & $<2.60\times10^{-5}$ & $0.2787_{-0.0125}^{+0.0110}$ & $70.56_{-1.17}^{+1.19}$ & $3.028_{-0.113}^{+0.110}$ & 0.85   \\
    FS + ($n$-free) & $<2.82\times10^{-5}$ & $0.2771\pm0.0114$ & $70.75_{-1.20}^{+1.22}$ & $2.990_{-0.123}^{+0.124}$ & 0.85 \\
    FS+BAO & $<2.52\times10^{-5}$ & $0.2947_{-0.0102}^{+0.0103}$ & $68.59_{-0.79}^{+0.81}$ & $3.065_{-0.107}^{+0.108}$ & 0.87 \\
    FS+BAO + (${n_{s}}$) & $<2.85\times10^{-5}$ & $0.2973\pm0.0083$ & $68.67_{-0.77}^{+0.76}$ & $3.044\pm0.090$ & 0.88 \\
    FS+BAO + (m.f.) & $<2.49\times10^{-5}$ & $0.2951\pm0.0101$ & $68.55\pm0.79$ & $3.021_{-0.095}^{+0.093}$ & 0.83 \\
    FS+BAO + ($k_\text{max}$) & $<1.69\times10^{-5}$ & $0.2806_{-0.0106}^{+0.0106}$ & $70.1050_{-1.0283}^{+1.0123}$ & $3.0219_{-0.1036}^{+0.1034}$ & 0.92 \\
    FS+BAO + CMB + $n_s$ 
    & $<2.50\times10^{-5}$ & $0.3066_{-0.0044}^{+0.0034}$ & $67.84_{-0.25}^{+0.33}$ & $3.058_{-0.080}^{+0.079}$ & 1.72 \\
    \hline
    \textbf{HS - $wowa$CDM} &  &  &  & & \\
    FS+BAO+CMB+DES-Y5  & $<2.76\times 10^{-5}$ & $0.3145_{-0.0063}^{+0.0065}$ & $67.13\pm0.64$ & $2.950\pm0.096$ & 1.95 \\
    FS+BAO+CMB+DES-Dovekie  & $<2.54\times 10^{-5}$ & $0.3092_{-0.0058}^{+0.0057}$ & $67.70_{-0.59}^{+0.61}$ & $2.906_{-0.087}^{+0.088}$ & 1.96 \\
    \hline
    \end{tabular}
    }
\caption{Parameter constraints from fits to the baseline models (denoted with gray bands) and their extensions. The top row is for GR, while the rest include the unscreened HS model with either $\Lambda$CDM or $wowa$ backgrounds. Full-Shape (FS) and BAO are from DESI Data Release 1 (DR1). We display 95\% upper bounds for $f_{R_0}$, while 68\% c.i. for the rest of the parameters. 
All models have $k_\text{max}=0.17 \hmpci$ in the FS analysis except for that labeled $(k_\text{max})$, which has $k_\text{max}=0.20\hmpci$. 
The CMB label refers to CMB$_{(\theta_*,\omega_{b},\omega_{bc})}$.
Models without CMB compressed or the label ($n_s$) contain the 10x Planck prior on $n_s$ (referred as ${n_{s,10}}$ in \cref{tab:priors}). The label (m.f.) stands for minimal freedom on $b_s$ and $n$-free is the full HS model without fixing $n=1$, where we find the value of $n=1.41\pm0.52$}.

\label{tab:results}
\end{table*}

\subsubsection*{Projection effects}

\begin{figure}[htbp]
    \centering
    \includegraphics[width=0.8 \textwidth]{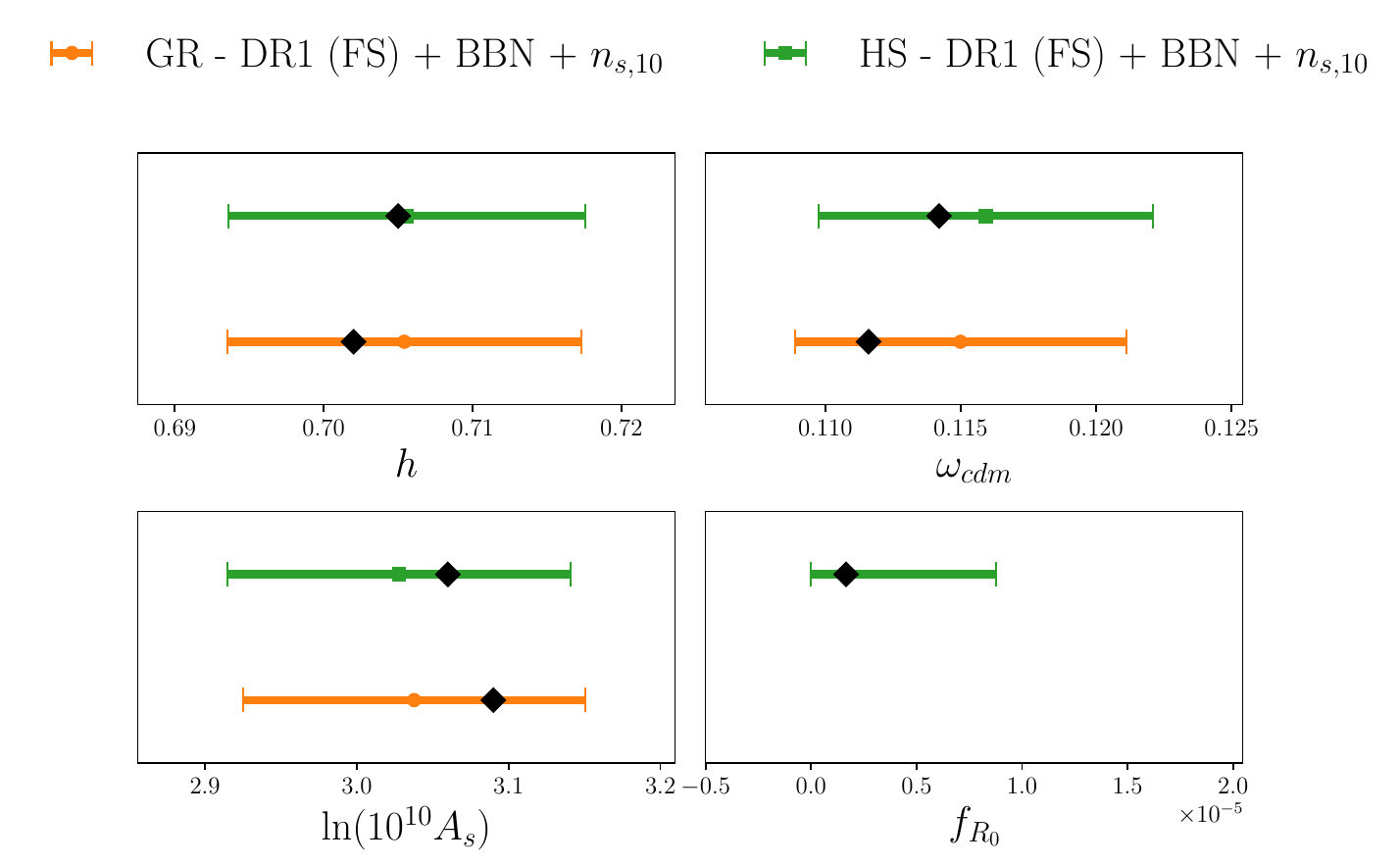}
    \caption{MAP values, denoted by black diamond markers, are shown for the varied cosmological parameters and $f_{R_0}$. The bars indicate the $1\sigma$ intervals of the projected posterior distributions obtained from the Full-Shape analyses. All MAP values lie within $0.5\sigma$ of the posterior best fits. }
    \label{fig:MAP}
\end{figure}

\begin{figure}[htbp]
    \centering
    \includegraphics[width=\linewidth]{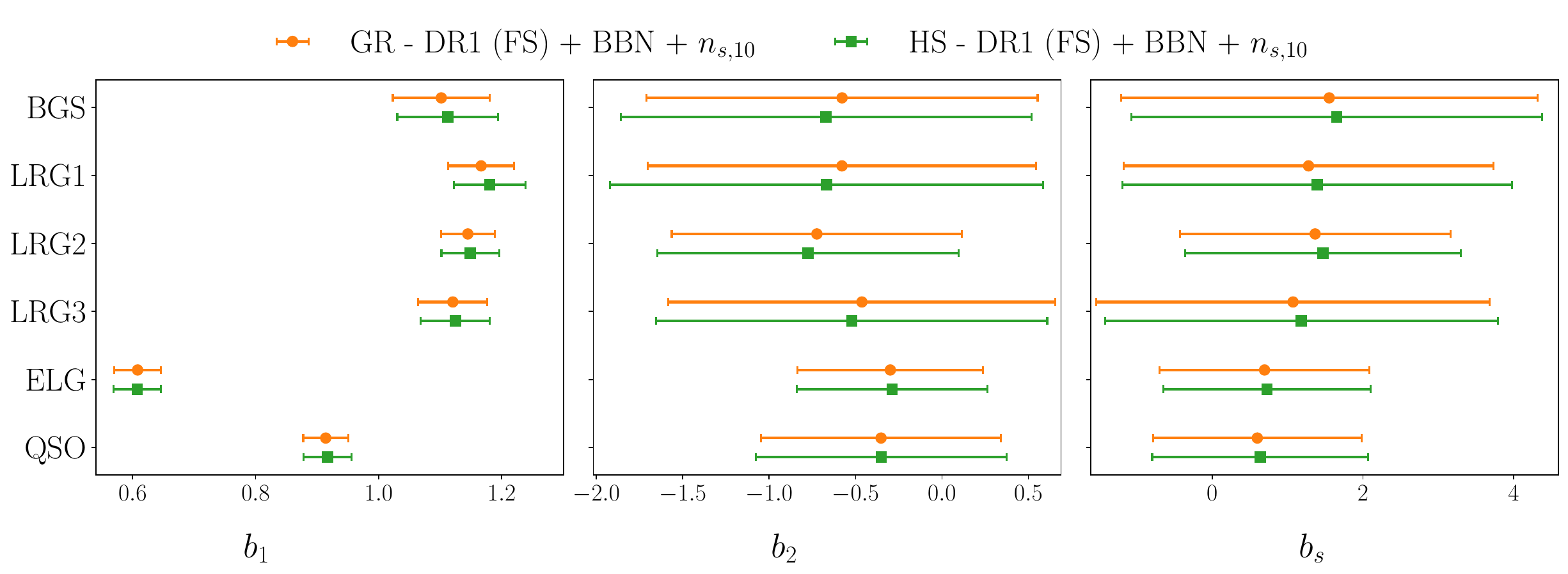}
    \caption{Whisker plots of the bias parameters for each tracer, comparing the GR and HS baseline models. Adding the additional degree of freedom from modified gravity, $f_{R_0}$, does not introduce degeneracy in the bias parameters, which remain consistent with GR.}
    \label{fig:bias_comparison}
\end{figure}

As discussed previously, FS analyses over finite-volume samples such as DESI may exhibit projection effects when nuisance parameters explore regions of parameter space that are not well constrained by the data. In principle, introducing additional model parameters could enhance such effects if they are degenerate with existing nuisance or cosmological parameters. In our case, however, the MG parameter $f_{R_0}$ appears to be only weakly correlated with the other parameters in the model. When comparing the HS and GR baseline models, the posterior distributions of the shared cosmological parameters remain essentially unchanged, indicating that $f_{R_0}$ is approximately orthogonal to the rest of the parameter space. This behavior is shown in  \cref{fig:bias_comparison}, where the projected bias parameters are consistent between the two models, and in \cref{figure:triangular_desiy1_fs}, where the cosmological parameter posteriors remain stable when including $f_{R_0}$. Overall, the analysis suggests that introducing $f_{R_0}$ in these MG models does not lead to new projection effects. 

 To test this statement quantitatively, we compute the MAP estimates using the \code{iminuit}\footnote{\href{https://github.com/scikit-hep/iminuit}{https://github.com/scikit-hep/iminuit}} package \cite{James:1975dr}. Projection effects are reflected in the discrepancies between the MAP values, shown as black diamond markers in \cref{fig:MAP}, and the maxima of the marginalized posterior distributions, shown as square markers. All MAP values lie within $0.5\sigma$ of the best-fit values of the marginalized posteriors, showing that the projection effects are small, if present at all.

\subsubsection*{Consistency and validation of MG baseline model}

\begin{figure}[htbp]
    \centering
    \includegraphics[width=0.8 \textwidth]{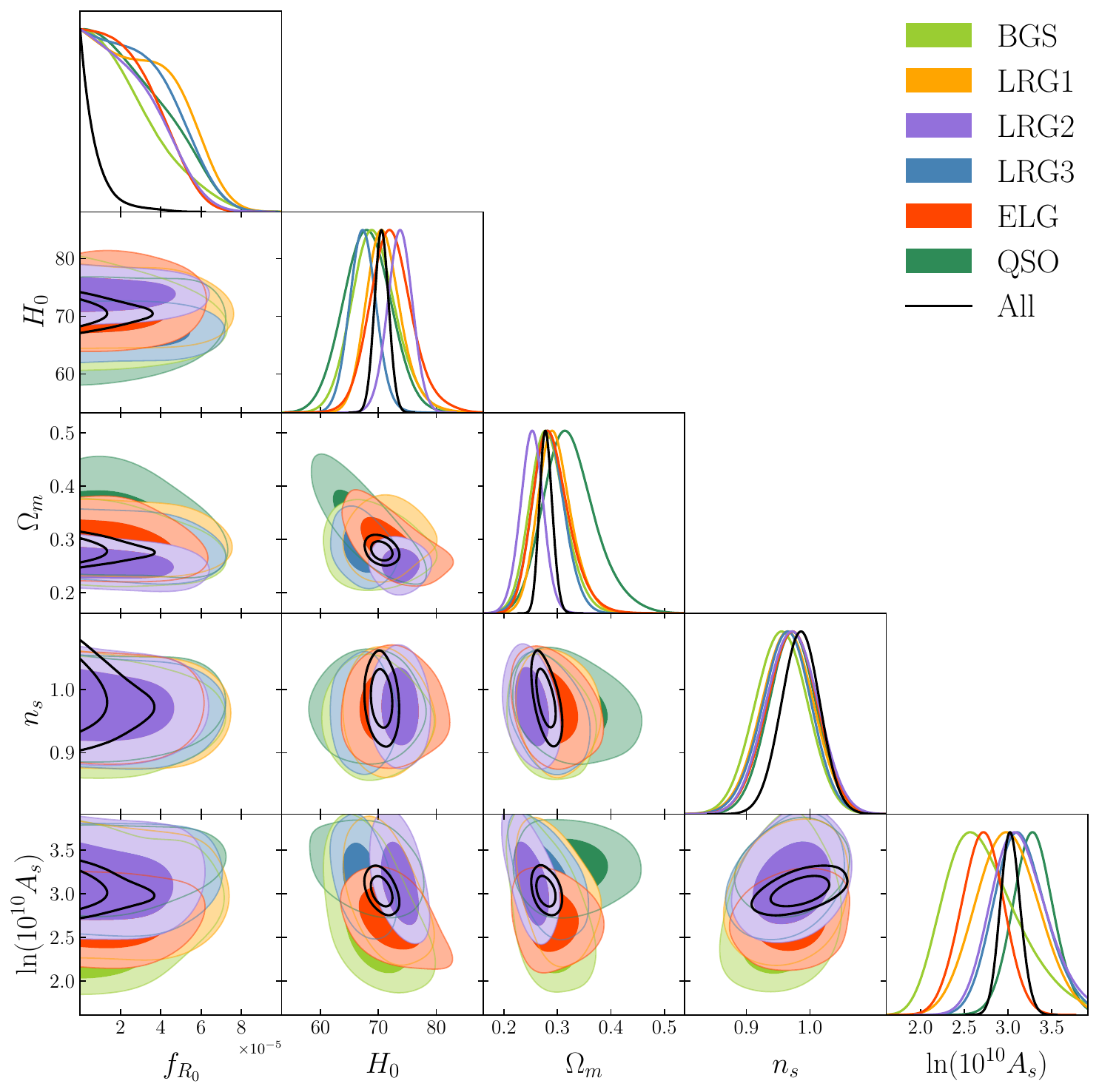}
    \caption{Posterior distributions of baseline model over individual DESI redshift bins. Results are self-consistent supporting the multi-bin analysis. }
    \label{fig:HS-FS_tracers}
\end{figure}

We run a few tests on the baseline model to explore its theoretical and observational consistency. First, we run the baseline model on each DESI redshift bin separately, finding that the inferred cosmologies are mutually consistent, as shown in \cref{fig:HS-FS_tracers}. Although the LRG1 bin presents a small bump in the posterior of $f_{R_0}$away from zero, this feature is not observed in the neighboring bins (BGS and LRG2) and does not show a consistent trend with redshift.

\begin{figure}[htbp]
    \centering
    \includegraphics[width=0.8\textwidth]{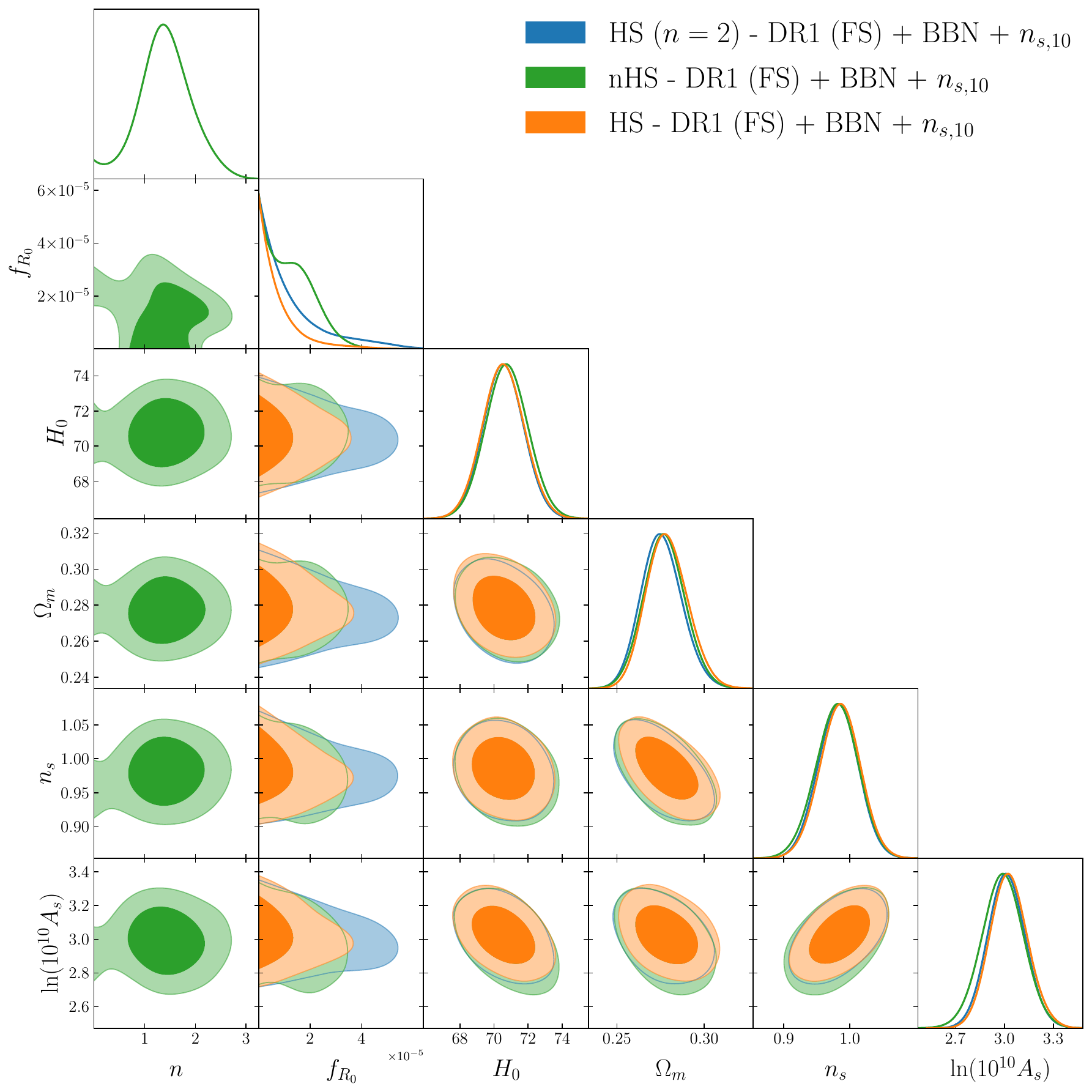}
    \caption{Full Shape analysis using the HS baseline model versus the full HS theory (nHS), which has $n$ free in equation \ref{eq:hu-sawicki-model}. Although the posterior distribution is in good agreement with the $n=1$ choice of the baseline model, there is a small degeneracy with $f_{R_0}$, which relaxes the bound on $m_\phi$ as shown in Figure \ref{fig:m_phi}.}
    \label{fig:nHS_constraint}
\end{figure}

In terms of theoretical consistency, we test the assumption of using $n=1$ in the HS model, \cref{eq:hu-sawicki-model}, by allowing $n$ to be free while keeping all other parameters and priors unchanged. We find $n=1.41\pm0.52$ over FS measurements only, while it shifts to $n=1.30 \pm 0.51$ when BAO is added. These results suggest that fixing $n=1$ is a consistent choice for the baseline model, and also that most information to constrain $n$ is in FS analysis. In addition, other parameters remain consistent with the choice $n=1$, as shown in Figure \ref{fig:nHS_constraint}. As also appreciated in the same figure, the small bump in the 1d posterior of $f_{R_0}$ about $n=2$ is due to the small degeneracy between these two parameters and not a signal of modified gravity, as can be seen when $n$ is initially fixed to 2.

\begin{figure}[htbp]
    \centering
    \includegraphics[width=0.6 \textwidth]{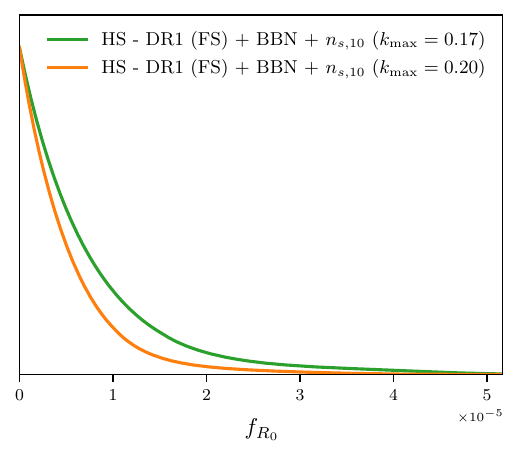}
    \caption{Comparison of the unscreened Hu-Sawicki (HS) baseline model inferred parameters using DESI standard choice of $k_{\mathrm{max}}=0.2$ \cite{DESI2024.V.KP5} in the FS analysis, instead of the more conservative value of $k_{\mathrm{max}}=0.17 \hmpci$ in our baseline, which avoids scales where the HS screening mechanism becomes relevant. } 
    \label{fig:kmax}
\end{figure}

The constraining power of higher wavenumbers on cosmological parameters increases until the inference becomes biased, as the model can no longer describe the non-linear matter clustering. An extensive study by the DESI team \cite{KP5s1-Maus,KP5s2-Maus,KP5s3-Noriega,KP5s4-Lai,KP5s5-Ramirez,DESI2024.V.KP5} using noiseless synthetic data and simulated data concluded that $k_\text{max}\sim 0.2 \hmpci$ is optimal for the DESI DR1 FS analysis using EFT. In the presence of an additional gravitational scalar sector, such as in $f(R)$ gravity, one would expect non-linearities in this sector to represent more important correction at $k_\text{max}=0.2 \hmpci$ than in GR. For the particular HS $f(R)$ model, these non-linear interactions are responsible for recovering GR at higher $k$, allowing the model to remain consistent with solar system gravitational measurements.
We conservatively adopt $k_\text{max}=0.17 \hmpci$ in order to test departures from GR on scales where the screening mechanism is well accounted for by the EFT counterterms \cite{Rodriguez-Meza:2023rga}. However, if one assumes that there is no screening mechanism at short scales, or that it remains degenerate with EFT parameters (which is expected for theories with $f_{R_0}$ of order $10^{-5}$ or smaller), then one may ask what additional constraining power is gained by using $k_\text{max}=0.2 \hmpci$, as in the $\Lambda$CDM case. The answer, shown in \cref{fig:kmax} and expected from  \cref{fig:mg_scale}, is that there is a $35\%$ improvement in the bounds of $f_{R_0}$ using the unscreened Hu-Sawicki model, as well as considerable improvements in the other cosmological parameters. However, it is clear from  \cref{fig:kmax} that most of the constraining power on the fifth-force scale comes from wavenumbers higher than $k_\text{max}=0.17 \hmpci$.
Moreover, in the case of $\Lambda$CDM, this cut-off scale can be pushed to slightly higher values, e.g. $k_\text{max}\sim 0.3 \hmpci$, by introducing a damping factor along the line of sight, as implemented in the \textsc{FolpsD} code \cite{folps-d}.\footnote{\url{https://github.com/alejandroaviles/folpsD}.} Although this phenomenological prescription can also be included in our MG model, the final constraining power is expected to be affected by degeneracies between the damping factor and the MG signal, as also occurs with massive neutrinos or noise-dominated samples \cite{folps-d}.

\section{MG and neutrino masses}
\label{sec:MGmnu}

The mass of neutrinos affects structure formation, and hence the distribution of galaxies, through two effects \cite{2012arXiv1212.6154L}. First, primordial neutrinos decouple from the primeval plasma while still relativistic at very early times, and they become nonrelativistic well after recombination. They therefore alter the background expansion relative to a universe with massless neutrinos. The second effect is related to their large velocity dispersion, which prevents neutrinos from clustering below a certain free-streaming scale, $\lambda_{\rm fs}$. As a result, the power spectrum exhibits a suppression at wavenumbers around $k_{\rm fs}=1/\lambda_{\rm fs}$ and above. While the most constraining effect comes from the expansion history of the Universe, as measured by combining BAO with early-time CMB information, DESI alone can also constrain the neutrino mass using only Full-Shape data, without external datasets, through the suppression of the power spectrum \cite{Y3.cpe-s2.Elbers.2025}.

On the other hand, one of the signatures of MG is an enhancement of the power spectrum above the scale $k_{\rm MG}$. It is then natural to ask how degenerate MG is with the neutrino mass. To explore this question, we include the sum of neutrino masses, $\sum m_\nu$, as an additional parameter, and an analysis using DESI DR1 (FS+BAO) + BBN + $n_{s,10} + \sum m_\nu$ shows the following constraints (at 95\% C.L.) on the sum of neutrino masses:
\begin{align}
&\text{GR:} \quad \sum m_\nu < 0.533 \,\text{eV} \qquad (95\% \,\text{c.l.}),\\[4pt]
&\text{MG:} \quad \sum m_\nu < 0.589 \,\text{eV} \qquad (95\% \,\text{c.l.}).
\end{align}
\begin{figure}[htbp]
    \centering
    \includegraphics[width=0.8 \textwidth]{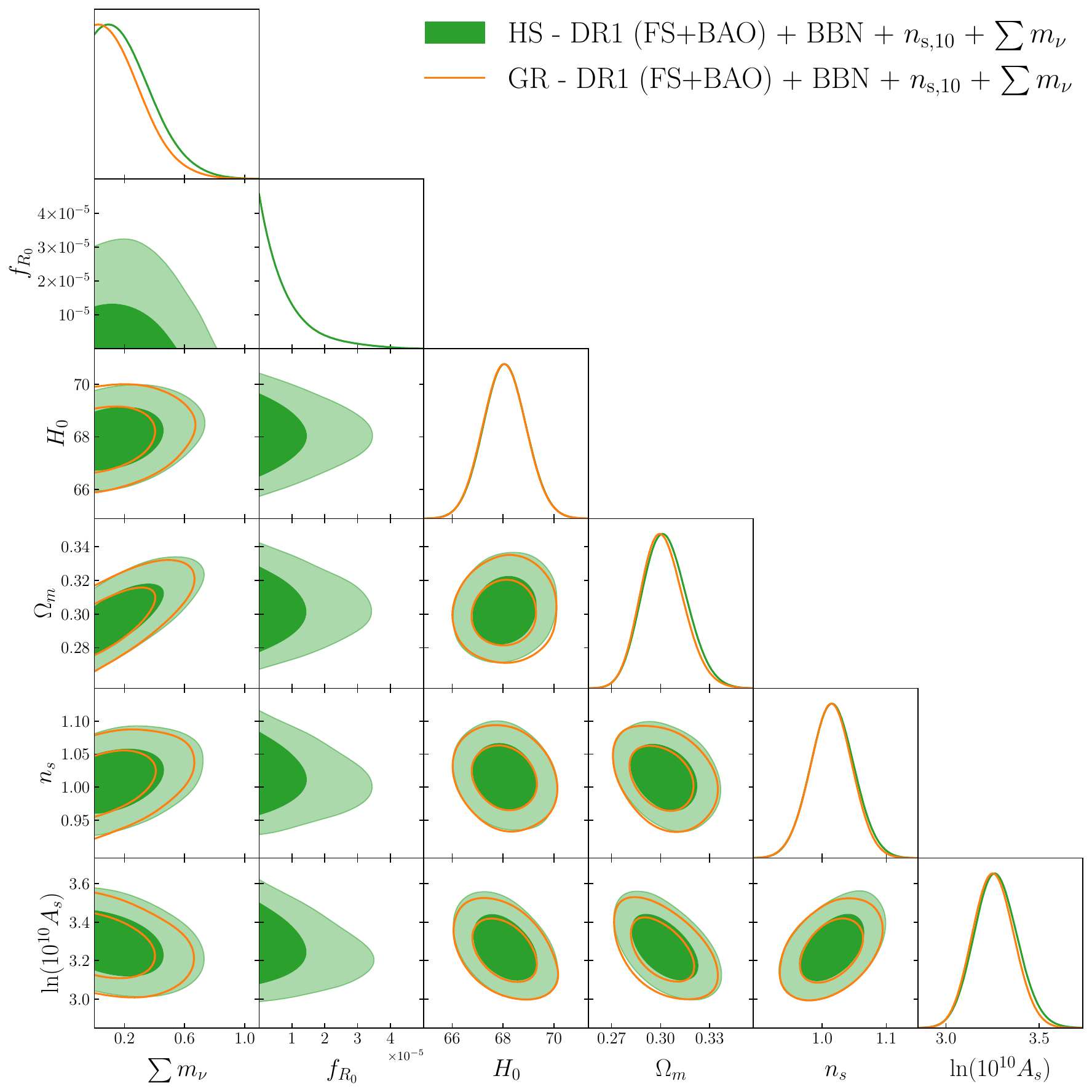}
    \caption{Effect of a free neutrino mass parameter in the cosmological inferred parameters for the MG with respect to GR. The degeneracy between $f_{R0}$ and $\sum m_\nu$ is certainly small for the HS model in the explored $k$-range for the FS analysis.} 
    \label{fig:mnu}
\end{figure}

The cosmological parameter posteriors for GR and the (unscreened) HS models in the presence of a free neutrinos mass parameter are shown in \cref{fig:mnu}. While we do observe a degradation in the neutrino-mass constraint when introducing the MG parameter $f_{R_0}$, this effect is small, at only about $10\%$. On the other hand, the constraint on $f_{R_0}$ is
\begin{align}
&f_{R_0} < 2.60 \times 10^{-5} (95\%  \,\text{c.l.}) \qquad \text{DESI DR1} \text{ (FS+BAO)+ BBN + }  n_{s,10} + \sum m_\nu.
\end{align}
This result should be contrasted with the bound reported in \cref{tab:results} for the same datasets that put an upper bound of $f_{R_0}<2.52\times 10^{-5}$ at 95\% level, which is only $3\%$ more constrictive than the result obtained in this section.

We conclude that, although there is a degeneracy between the neutrino mass and the parameter $f_{R_0}$ in the unscreened HS model, it is small.

\section{Final thoughts}
\label{sec:conclusions}

Galaxy clustering is one of the most powerful probes to constrain scale-dependent modifications to gravity. Spectroscopic galaxy surveys, in particular, provide clean and tight bounds, since these models mainly affect the growth of structure at late times and introduce characteristic scales in the matter distribution. This limits the sensitivity of other probes: the CMB mostly traces early-time physics, while background observables such as SNIa or BAO probe the expansion history but are largely insensitive to scale-dependent effects. Other probes can also be affected by additional uncertainties, such as strong baryonic effects or poorly understood physics. In contrast, measurements of the FS galaxy power spectrum directly probe these signatures. Further, they do not introduce new prior-volume effects beyond those already present in $\Lambda$CDM. This makes surveys like DESI especially well suited to constrain these models on their own, without relying on external datasets.

In this work we have investigated the ability of galaxy clustering measurements to constrain scale-dependent modifications of gravity using the DESI DR1 FS power spectrum. We obtain constraints on the HS model ranging from $f_{R_0} < 2.60 \times 10^{-5}\;$ to $\;f_{R_0} <1.69 \times 10^{-5}$ (95\% c.l.), depending on the datasets and scale cuts used in the analyses. In MG scenarios that aim to explain cosmic acceleration without dark energy, deviations from GR arise below a characteristic physical scale while large-scale clustering remains essentially unchanged. This introduces a new scale in the growth of structure, $k_\text{MG}$, which enhances the galaxy power spectrum for $k \gtrsim k_\text{MG}$ while recovering the standard GR prediction on larger scales. Our bounds translate into a physical range for the fifth force associated with these MG models of $\lambda < 17.81 \,\text{Mpc}$. Since these scales are close to the limits typically used in spectroscopic survey EFT analyses ($k \lesssim 0.20 \hmpci$), it is unlikely to obtain substantial improvements using only the galaxy power spectrum. However, DESI DR2 and subsequent releases will include a larger survey volume and improved statistics that can survey larger $k$ modes, potentially improving constraints on the characteristic MG scale and the associated parameters. In addition, higher-order clustering statistics such as the galaxy bispectrum provide complementary information on the growth of structure and non-linear gravitational dynamics. Including the galaxy bispectrum is expected to further enhance the constraining power on MG models.

Finally, important synergies are expected with upcoming photometric galaxy surveys. Wide photometric surveys probe larger cosmological volumes and higher redshifts, while spectroscopic surveys such as DESI provide precise three-dimensional clustering measurements. The combination of these datasets will enable improved tests of gravity across a wide range of scales and cosmic times, providing a powerful avenue to search for possible deviations from GR.

\section{Data Availability}

Data from the plots in this paper are available on Zenodo as part of DESI's Data Management
Plan (\url{https://doi.org/10.5281/zenodo.19828033}). The data used in this analysis is public along the Data Release 1 (details in \url{https://data.desi.lbl.gov/doc/releases/dr1/}).


\acknowledgments

DG, GN, AA and HN are supported by SECIHTI grant CBF2023-2024-162. AA and HN are supported by DGAPA-PAPIIT IA101825. AA also acknowledges  DGAPA-PAPIIT IG102123. GN, AA and DG acknowledge financial support from SECIHTI grant CBF-2025-I-2795 and DAIP-UG grant CIIC-254/2026. GN and DG also thank the DCI-UG DataLab for computational resources. CGQ acknowledges support provided by NASA through the NASA Hubble Fellowship grant HST-HF2-51554.001-A awarded by the Space Telescope Science Institute, which is operated by the Association of Universities for Research in Astronomy, Inc., for NASA, under contract NAS5-26555.

This material is based upon work supported by the U.S. Department of Energy (DOE), Office of Science, Office of High-Energy Physics, under Contract No. DE–AC02–05CH11231, and by the National Energy Research Scientific Computing Center, a DOE Office of Science User Facility under the same contract. Additional support for DESI was provided by the U.S. National Science Foundation (NSF), Division of Astronomical Sciences under Contract No. AST-0950945 to the NSF’s National Optical-Infrared Astronomy Research Laboratory; the Science and Technology Facilities Council of the United Kingdom; the Gordon and Betty Moore Foundation; the Heising-Simons Foundation; the French Alternative Energies and Atomic Energy Commission (CEA); the National Council of Humanities, Science and Technology of Mexico (CONAHCYT); the Ministry of Science, Innovation and Universities of Spain (MICIU/AEI/10.13039/501100011033), and by the DESI Member Institutions: \url{https://www.desi.lbl.gov/collaborating-institutions}. Any opinions, findings, and conclusions or recommendations expressed in this material are those of the author(s) and do not necessarily reflect the views of the U. S. National Science Foundation, the U. S. Department of Energy, or any of the listed funding agencies.

The authors are honored to be permitted to conduct scientific research on I'oligam Du'ag (Kitt Peak), a mountain with particular significance to the Tohono O’odham Nation.


\bibliographystyle{JHEP}
\bibliography{refs,DESI_supporting_papers}


\appendix


\section{Author Affiliations}
\label{sec:affiliations}

{\small

\noindent \hangindent=.5cm $^{1}${Departamento de F\'{\i}sica, DCI-Campus Le\'{o}n, Universidad de Guanajuato, Loma del Bosque 103, Le\'{o}n, Guanajuato C.~P.~37150, M\'{e}xico}

\noindent \hangindent=.5cm $^{2}${Instituto de Ciencias F\'{\i}sicas, Universidad Nacional Aut\'onoma de M\'exico, Av. Universidad s/n, Cuernavaca, Morelos, C.~P.~62210, M\'exico}

\noindent \hangindent=.5cm $^{3}${Instituto Avanzado de Cosmolog\'{\i}a A.~C., San Marcos 11 - Atenas 202. Magdalena Contreras. Ciudad de M\'{e}xico C.~P.~10720, M\'{e}xico}

\noindent \hangindent=.5cm $^{4}${Center for Astrophysics $|$ Harvard \& Smithsonian, 60 Garden Street, Cambridge, MA 02138, USA}

\noindent \hangindent=.5cm $^{5}${NASA Einstein Fellow}

\noindent \hangindent=.5cm $^{6}${Lawrence Berkeley National Laboratory, 1 Cyclotron Road, Berkeley, CA 94720, USA}

\noindent \hangindent=.5cm $^{7}${Department of Physics, Boston University, 590 Commonwealth Avenue, Boston, MA 02215 USA}

\noindent \hangindent=.5cm $^{8}${Dipartimento di Fisica ``Aldo Pontremoli'', Universit\`a degli Studi di Milano, Via Celoria 16, I-20133 Milano, Italy}

\noindent \hangindent=.5cm $^{9}${INAF-Osservatorio Astronomico di Brera, Via Brera 28, 20122 Milano, Italy}

\noindent \hangindent=.5cm $^{10}${Department of Physics \& Astronomy, University College London, Gower Street, London, WC1E 6BT, UK}

\noindent \hangindent=.5cm $^{11}${Instituto de F\'{\i}sica, Universidad Nacional Aut\'{o}noma de M\'{e}xico,  Circuito de la Investigaci\'{o}n Cient\'{\i}fica, Ciudad Universitaria, Cd. de M\'{e}xico  C.~P.~04510,  M\'{e}xico}

\noindent \hangindent=.5cm $^{12}${IRFU, CEA, Universit\'{e} Paris-Saclay, F-91191 Gif-sur-Yvette, France}

\noindent \hangindent=.5cm $^{13}${University of California, Berkeley, 110 Sproul Hall \#5800 Berkeley, CA 94720, USA}

\noindent \hangindent=.5cm $^{14}${Departamento de F\'isica, Universidad de los Andes, Cra. 1 No. 18A-10, Edificio Ip, CP 111711, Bogot\'a, Colombia}

\noindent \hangindent=.5cm $^{15}${Observatorio Astron\'omico, Universidad de los Andes, Cra. 1 No. 18A-10, Edificio H, CP 111711 Bogot\'a, Colombia}

\noindent \hangindent=.5cm $^{16}${Institut d'Estudis Espacials de Catalunya (IEEC), c/ Esteve Terradas 1, Edifici RDIT, Campus PMT-UPC, 08860 Castelldefels, Spain}

\noindent \hangindent=.5cm $^{17}${Institute of Cosmology and Gravitation, University of Portsmouth, Dennis Sciama Building, Portsmouth, PO1 3FX, UK}

\noindent \hangindent=.5cm $^{18}${Institute of Space Sciences, ICE-CSIC, Campus UAB, Carrer de Can Magrans s/n, 08913 Bellaterra, Barcelona, Spain}

\noindent \hangindent=.5cm $^{19}${University of Virginia, Department of Astronomy, Charlottesville, VA 22904, USA}

\noindent \hangindent=.5cm $^{20}${Fermi National Accelerator Laboratory, PO Box 500, Batavia, IL 60510, USA}

\noindent \hangindent=.5cm $^{21}${Department of Astronomy, University of Texas at Austin, 2515 Speedway, TX 78712, USA}

\noindent \hangindent=.5cm $^{22}${Center for Cosmology and AstroParticle Physics, The Ohio State University, 191 West Woodruff Avenue, Columbus, OH 43210, USA}

\noindent \hangindent=.5cm $^{23}${Department of Physics, The Ohio State University, 191 West Woodruff Avenue, Columbus, OH 43210, USA}

\noindent \hangindent=.5cm $^{24}${The Ohio State University, Columbus, 43210 OH, USA}

\noindent \hangindent=.5cm $^{25}${Department of Physics, University of Michigan, 450 Church Street, Ann Arbor, MI 48109, USA}

\noindent \hangindent=.5cm $^{26}${University of Michigan, 500 S. State Street, Ann Arbor, MI 48109, USA}

\noindent \hangindent=.5cm $^{27}${Department of Physics, The University of Texas at Dallas, 800 W. Campbell Rd., Richardson, TX 75080, USA}

\noindent \hangindent=.5cm $^{28}${NSF NOIRLab, 950 N. Cherry Ave., Tucson, AZ 85719, USA}

\noindent \hangindent=.5cm $^{29}${Department of Physics, Southern Methodist University, 3215 Daniel Avenue, Dallas, TX 75275, USA}

\noindent \hangindent=.5cm $^{30}${Department of Physics and Astronomy, University of California, Irvine, 92697, USA}

\noindent \hangindent=.5cm $^{31}${Sorbonne Universit\'{e}, CNRS/IN2P3, Laboratoire de Physique Nucl\'{e}aire et de Hautes Energies (LPNHE), FR-75005 Paris, France}

\noindent \hangindent=.5cm $^{32}${Departament de F\'{i}sica, Serra H\'{u}nter, Universitat Aut\`{o}noma de Barcelona, 08193 Bellaterra (Barcelona), Spain}

\noindent \hangindent=.5cm $^{33}${Institut de F\'{i}sica d’Altes Energies (IFAE), The Barcelona Institute of Science and Technology, Edifici Cn, Campus UAB, 08193, Bellaterra (Barcelona), Spain}

\noindent \hangindent=.5cm $^{34}${Instituci\'{o} Catalana de Recerca i Estudis Avan\c{c}ats, Passeig de Llu\'{\i}s Companys, 23, 08010 Barcelona, Spain}

\noindent \hangindent=.5cm $^{35}${Department of Physics and Astronomy, University of Waterloo, 200 University Ave W, Waterloo, ON N2L 3G1, Canada}

\noindent \hangindent=.5cm $^{36}${Perimeter Institute for Theoretical Physics, 31 Caroline St. North, Waterloo, ON N2L 2Y5, Canada}

\noindent \hangindent=.5cm $^{37}${Waterloo Centre for Astrophysics, University of Waterloo, 200 University Ave W, Waterloo, ON N2L 3G1, Canada}

\noindent \hangindent=.5cm $^{38}${Departament de F\'isica, EEBE, Universitat Polit\`ecnica de Catalunya, c/Eduard Maristany 10, 08930 Barcelona, Spain}

\noindent \hangindent=.5cm $^{39}${Department of Physics and Astronomy, Sejong University, 209 Neungdong-ro, Gwangjin-gu, Seoul 05006, Republic of Korea}

\noindent \hangindent=.5cm $^{40}${Abastumani Astrophysical Observatory, Tbilisi, GE-0179, Georgia}

\noindent \hangindent=.5cm $^{41}${Department of Physics, Kansas State University, 116 Cardwell Hall, Manhattan, KS 66506, USA}

\noindent \hangindent=.5cm $^{42}${Faculty of Natural Sciences and Medicine, Ilia State University, 0194 Tbilisi, Georgia}

\noindent \hangindent=.5cm $^{43}${CIEMAT, Avenida Complutense 40, E-28040 Madrid, Spain}

\noindent \hangindent=.5cm $^{44}${Department of Physics \& Astronomy, Ohio University, 139 University Terrace, Athens, OH 45701, USA}
}

\end{document}